\documentclass[aps,prb,superscriptaddress,reprint]{revtex4-1}

\usepackage{graphicx}
\usepackage{subfigure}
\usepackage{caption}
\usepackage{epstopdf}
\usepackage{epsfig}

\usepackage{amssymb}
\usepackage{amsthm}

\begin{document}


\title{Evidence of quantum Griffiths phase in Ni$_{1-x}$V$_x$ nanoalloys}


\author{P. Swain}
\email[]{priya@phy.iitkgp.ernet.in}
\affiliation{Department of Physics, Indian Institute of Technology
Kharagpur, Kharagpur-721302, INDIA}
\author{Suneel K. Srivastava}
\affiliation{Department of Chemistry, Indian Institute of Technology
Kharagpur, Kharagpur-721302, INDIA}
\author{Sanjeev K. Srivastava}
\email[]{sanjeev@phy.iitkgp.ernet.in}
\affiliation{Department of Physics, Indian Institute of Technology
Kharagpur, Kharagpur-721302, INDIA}



\date{\today}

\begin{abstract}

Metallic Ni$_{1-x}$V$_x$ alloys are known to exhibit a ferromagnetic to paramagnetic disordered quantum phase transition (QPT) at the critical concentration $x_c \sim$ 0.114 in bulk. Such a QPT is accompanied by a quantum Griffiths phase (QGP), the physical observables in which follow non-universal power-law temperature dependences, in a finite temperature range on the paramagnetic side of the transition. In the present work, we explore the occurrence of QGP in nanoparticles of this alloy system. Nanoalloys with $x$ in the neighbourhood of $x_c$ and mean diameter 18-33 nm were prepared by a chemical reflux method. Following a few microscopic and spectroscopic studies to determine the sizes, compositions and phases, dc magnetization measurements were also performed to seek out any signature of QGP in the nanoalloys. A paramagnetic-like increase of magnetization is observed to emerge below an $x$-dependent transition temperature $\rm{T_P} (\it{x})$ within the blocked ferromagnetic state of the nanoparticles, and is corroborated by a peak at $\rm{T_P} (\it{x})$ in the temperature dependence of resistivity. The magnetic susceptibility in this emergent phase follows a non-Curie power-law temperature dependence below 10 K for $0.09 \leq x \leq 0.14$, indicating the presence of a QGP in the nanoparticles within these temperature and composition ranges.

\end{abstract}

%
%
%
\maketitle
%
%
\section{Introduction}

The ferromagnetic (FM) to paramagnetic (PM) transition temperature of the elemental ferromagnet Ni is known to be suppressible to absolute zero on alloying it with a critical concentration ($x_c$) of a non-magnetic $d$-element, like Pd,\cite{Nicklas99} Pt,\cite{Ododo77} Rh,\cite{Muellner} or V.\cite{Schroeder10} The ground state of such an alloy system, thus, undergoes an FM - PM quantum phase transition (QPT) across $x_c$, known as the quantum critical point (QCP). Among the Ni binary alloys, Ni$_x$Pd$_{1-x}$
and Ni$_{1-x}$V$_x$ have experimentally been shown to undergo a QPT.\cite{Nicklas99, Schroeder10} A QPT is driven by quantum fluctuations with an energy scale E$_Q$, which compete with thermal fluctuations of energy scale $k_B$T at finite temperatures, and dominate the system's properties over the latter for E$_Q > k_B$T.\cite{subir} As a consequence, the system exhibits unconventional physical behaviour, like a non-Fermi liquid (NFL) phase, characterized by non-universal power-law temperature dependences of physical observables, around QCP.\cite{subir} In case of Ni$_x$Pd$_{1-x}$, the compositional disorder at the QCP ($x_c~\sim~0.026$) is small, and hence the NFL behaviour is observable in this system.\cite{Nicklas99} The Ni$_{1-x}$V$_x$ system, however, is associated with a considerable compositional disorder at the QCP ($x_c~\sim~0.114$)\cite{Schroeder10} because of the larger $x_c$. The non-universal behaviour in such disordered systems is not limited just to the QCP region; rather, it extends in a finite temperature range, identifiable as a quantum Griffiths phase (QGP), on the paramagnetic side of the transition.\cite{Schroeder10, Schroeder11} At very low temperatures on the paramagnetic side, however, the non-universal behaviour is masked by the appearance of a cluster glass phase.

Fundamentally, a QPT in metals is proposed to be associated with a qualitative change in the Fermi surface (FS) in the vicinity of the QCP.\cite{sachdev11} This proposition can have important consequences in case of nanomaterials, wherein the quantum confinement effects lead to properties different from their bulk counterparts. Further, the FS of a nanoparticle is also supposedly different from the corresponding bulk FS,\cite{weismann, swain15} and may modify the quantum critical behaviour. This had led the authors earlier to investigate the occurrence of QPT in Ni$_x$Pd$_{1-x}$ nanoalloys,\cite{swain15} wherein the nanoalloys were found to exhibit a QPT, in spite of not showing any NFL behaviour. Along the same line, it is quite intriguing to investigate whether the QGP, the characteristic feature of the Ni$_{1-x}$V$_x$ bulk alloys, appears also in the nanoparticles of this alloy system, although the magnetic phase diagram of a nanoparticle system, which may include superparamgnetic (SPM), blocked FM, spin-glass, etc. phases, is more complex than the corresponding bulk,\cite{Hiroi} and may render the investigation difficult.

Ni nanoparticles\cite{Kazan, He, Fonesca, Senapati} and Ni-V alloy microparticles\cite{Bambhaniya13} have earlier been synthesized and shown to possess FM,\cite{Kazan, He} SPM,\cite{Fonesca} photocatalytic\cite{Senapati} and H-storage properties.\cite{Bambhaniya13} Nanoparticles of Ni-V alloys, on the other hand, have hitherto not been synthesized or studied for magnetic properties, or even for any application, to the best of the authors' knowledge. So, any kind of investigation, including the exploration of any signature of a QGP, on this nanoalloy system requires, as a pre-step, finding a method to prepare these nanoalloys.

In this work, we aim at exploring the signature of QGP in Ni$_{1-x}$V$_x$ nanoalloy system. For preparation of the nanoalloys, the chemical reflux method used to synthesize Ni$_x$Pd$_{1-x}$ nanoalloys in our previous work,\cite{swain15} but with an appropriately modified set of chemicals, was adopted. Nanoparticles of Ni, V and Ni$_{1-x}$V$_x$, with $x$ in the vicinity of $x_c$, were prepared this way for the investigations. After determining the sizes, phases and compositions by different microscopic and spectroscopic techniques, the existence of QGP was explored using dc magnetization and electrical resistivity measurements.

\section{Experimental Section}

The nanoparticles of Ni, V and Ni$_{1-x}$V$_x$ (0.05 $\leq x \leq$ 0.20) were synthesized basically by reduction of metal precursor salts vanadium (III) chloride (VCl$_3$.H$_2$O) and nickel (II) chloride (NiCl$_{2}$), either separately for the elemental cases or simultaneously with appropriate stoichiometry for the nanoalloys, by hydrazine hydrate in the presence of the surfactant diethanolamine in a conventional reflux apparatus.\cite{swain15} In the cases of elemental nanoparticles, typically 0.5 mmol of VCl$_3$ (NiCl$_{2}$) was dissolved in 30 ml distilled water to yield complexes of V$^{2+}$ (Ni$^{2+}$) ions in the solution; for nanoalloys, proportionately appropriate amounts of the two salts were dissolved sequentially in distilled water. Subsequently, 5 ml of diethanolamine was added as a surfactant to the above solution, followed by 6 ml of hydrazine hydrate as the common reducing agent. Finally, 40 ml distilled water was added to this, and the resulting solution was refluxed for 8 h at 110 $^{\circ}$C in an oil bath. The black-colored precipitate, i.e., the alloy, was then washed with warm distilled water, centrifuged at 3500 RPM and dried in vacuum for 48 h.

The morphologies of the nanoalloys were investigated using (i) a ZEISS SUPRA 40 field-emission scanning electron microscope and (ii) a JEOL JEM-2100 high resolution transmission electron microscope operated at 200 kV. A drop of the colloidal nanoparticles, pre-sonicated in acetone, was placed on a small quartz peace to prepare the sample for field-emission scanning electron microscopy (FESEM); the drops were placed on a carbon supported Cu transmission electron microscope grid for high resolution transmission electron microscopy (HRTEM) and selected area electron diffraction (SAED). Energy dispersive X-ray analyses (EDAX) of the nanoalloys were performed using a JEOL scanning electron microscope to determine the final synthesized composition $x_s$. The phases were determined by X-ray diffraction (XRD) on a Philips X'Pert MRD system using Cu K$_\alpha$ radiation operated at 45 kV and 40 mA. The stoichiometries of the samples were further studied using X-ray photoelectron spectroscopy (XPS). XPS spectra were recorded on a PHI 5000 Versaprobe II system using a micro-focused monochromatic Al $K _\alpha$ source ($h\nu$ = 1486.6 eV), a hemispherical analyser, and a multichannel detector. Charge neutralization in each measurement was achieved using a combination of low energy Ar$^+$ ions and electrons. The binding energy scale was charge referenced to C 1s peak at 284.5 eV. High-resolution XPS spectra were acquired at 58.7 eV analyzer pass energy in steps of 0.25 eV. Further, the temperature dependences of sample resistivities in the temperature range 15 K - 300 K were measured on pelletized nanoalloys by four-probe technique using a Lakeshore Resistivity 7500 set-up with the help of a nanovoltmeter as a current source. Finally, DC magnetizations versus temperature (5 K $\leq$ T $\leq$ 300 K) at 500 Oe and versus field (-5 T $\leq$ H $\leq$ 5 T) at 2 K were measured using either a cryogen-free 9 T CRYOGENIC physical property measurement system or a Quantum Design MPMS SQUID VSM EverCool system.

\section{Results and Discussion}

\subsection{EDAX}
\begin{figure}%
\centering
\includegraphics[width=0.5\textwidth]{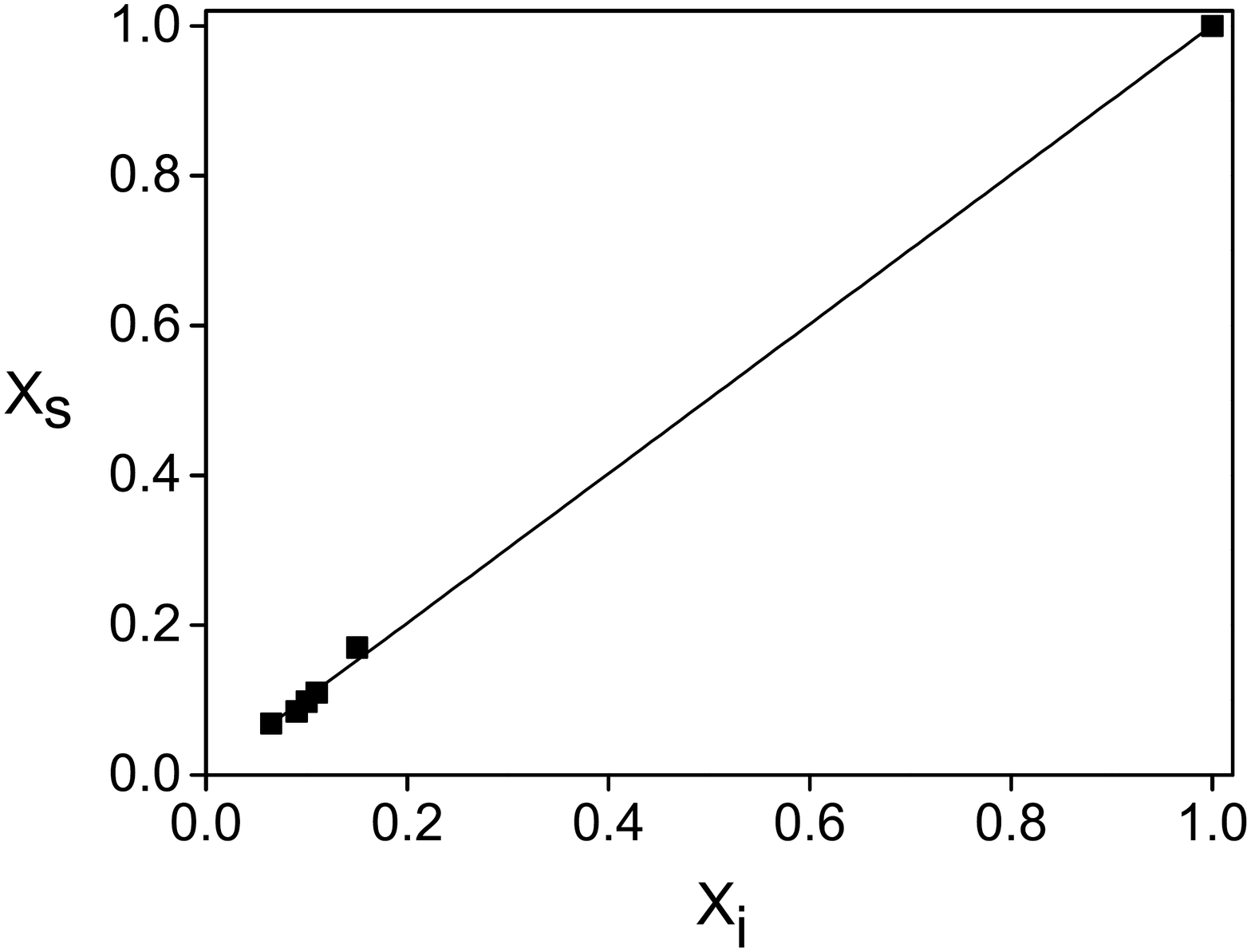}%
\caption{Variation of the composition $x_s$ determined from EDAX
and the initial composition $x_i$}
\end{figure}

The variation of EDAX determined composition $x_s$ with the initial composition $x_i$ is plotted in Fig. 1. As is clearly evident from the figure, $x_s$ shows a linear variation ($x_s$ = 0.0034 + 0.997 $x_i$) with $x_i$ and confirms that the stoichiometries taken during the syntheses are essentially the same as in the finally synthesized samples. Henceforth, the value of $x$ in Ni$_{1-x}$V$_x$ will be taken as $x_s$.

\subsection{FESEM and HRTEM}

\begin{figure}%
\centering
\subfigure[][]{%
\includegraphics[width=0.20\textwidth]{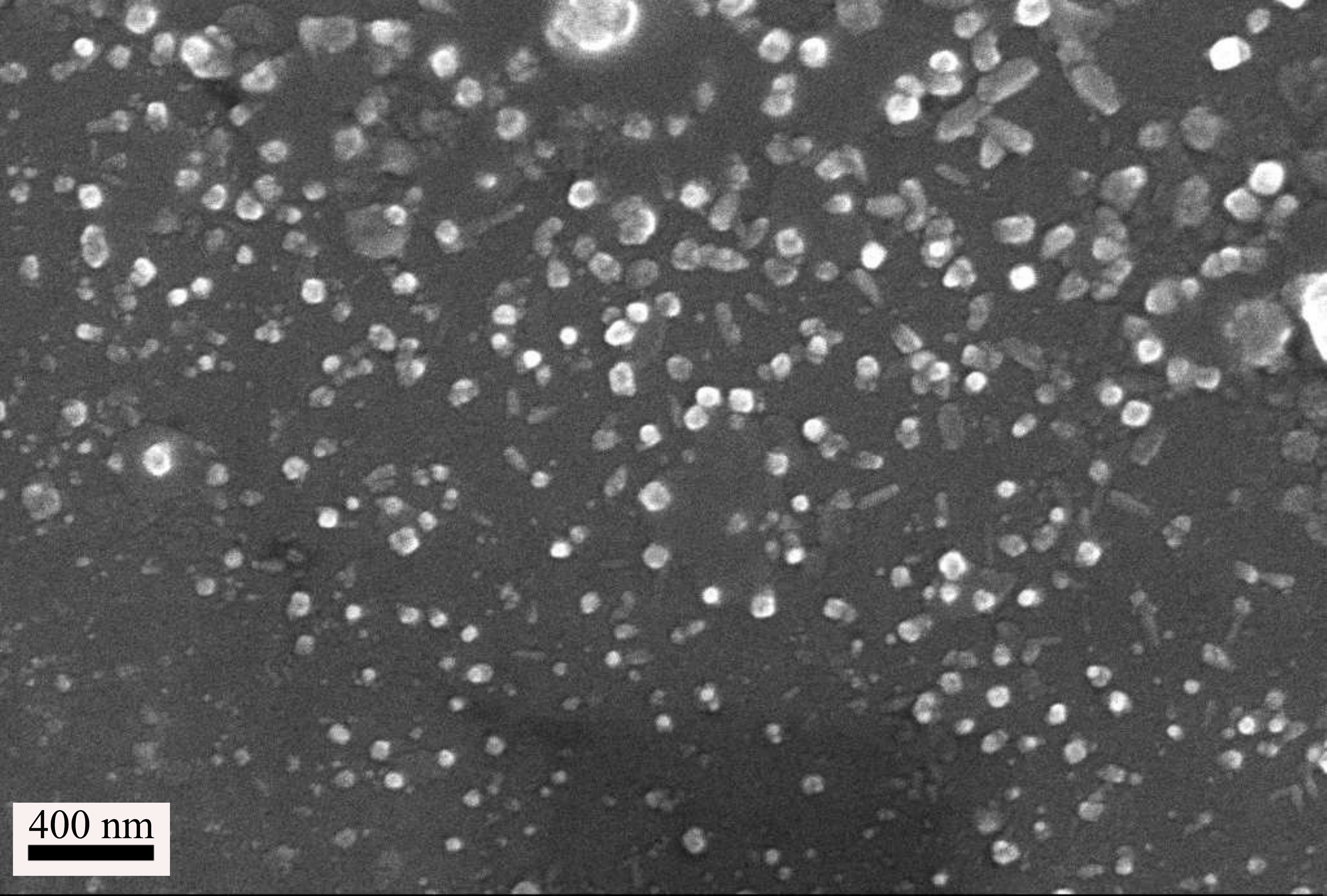}}%
\hspace{8pt}%
\subfigure[][]{%
\includegraphics[width=0.22\textwidth]{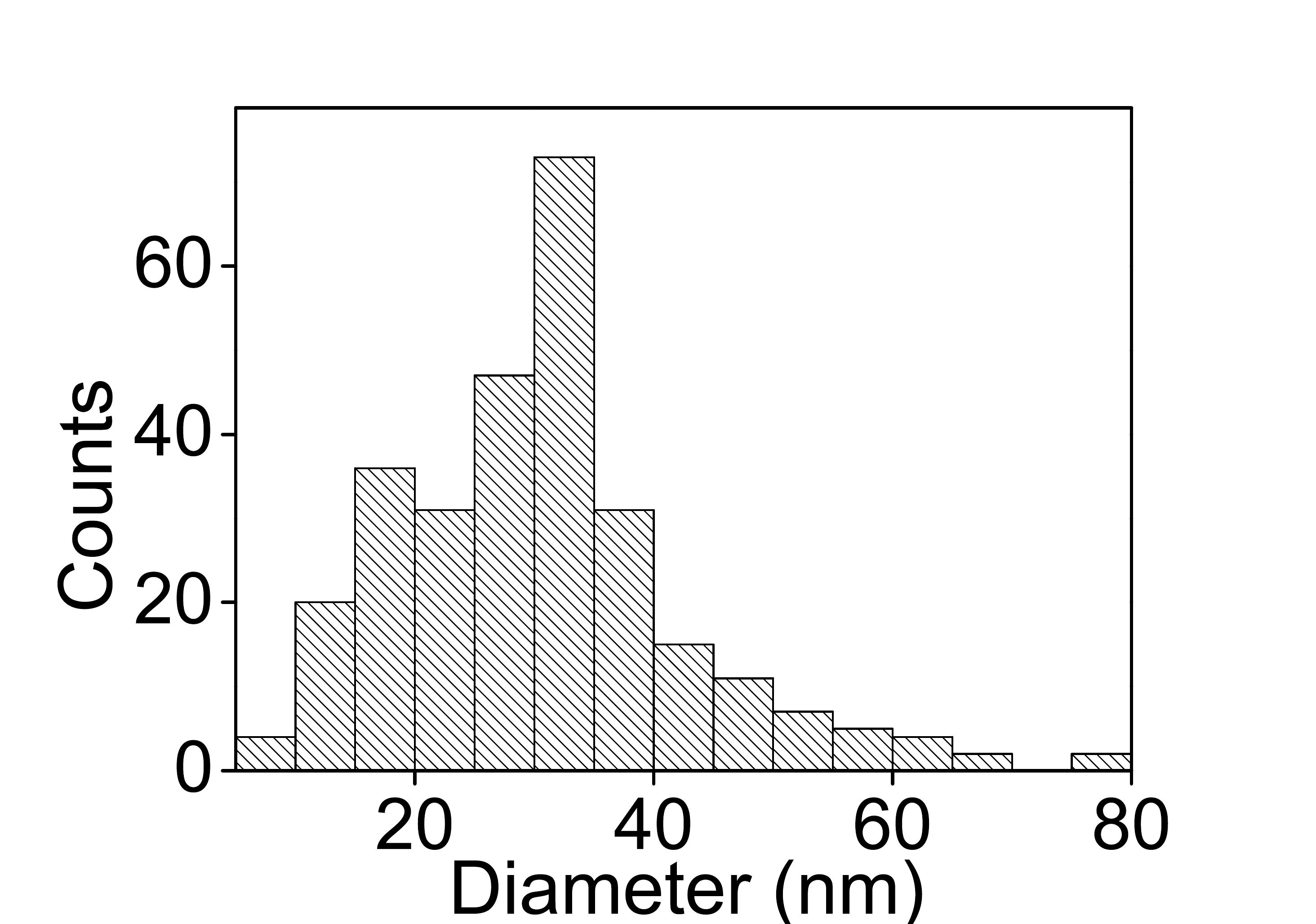}}%
\hspace{8pt}%
\subfigure[][]{%
\includegraphics[width=0.20\textwidth]{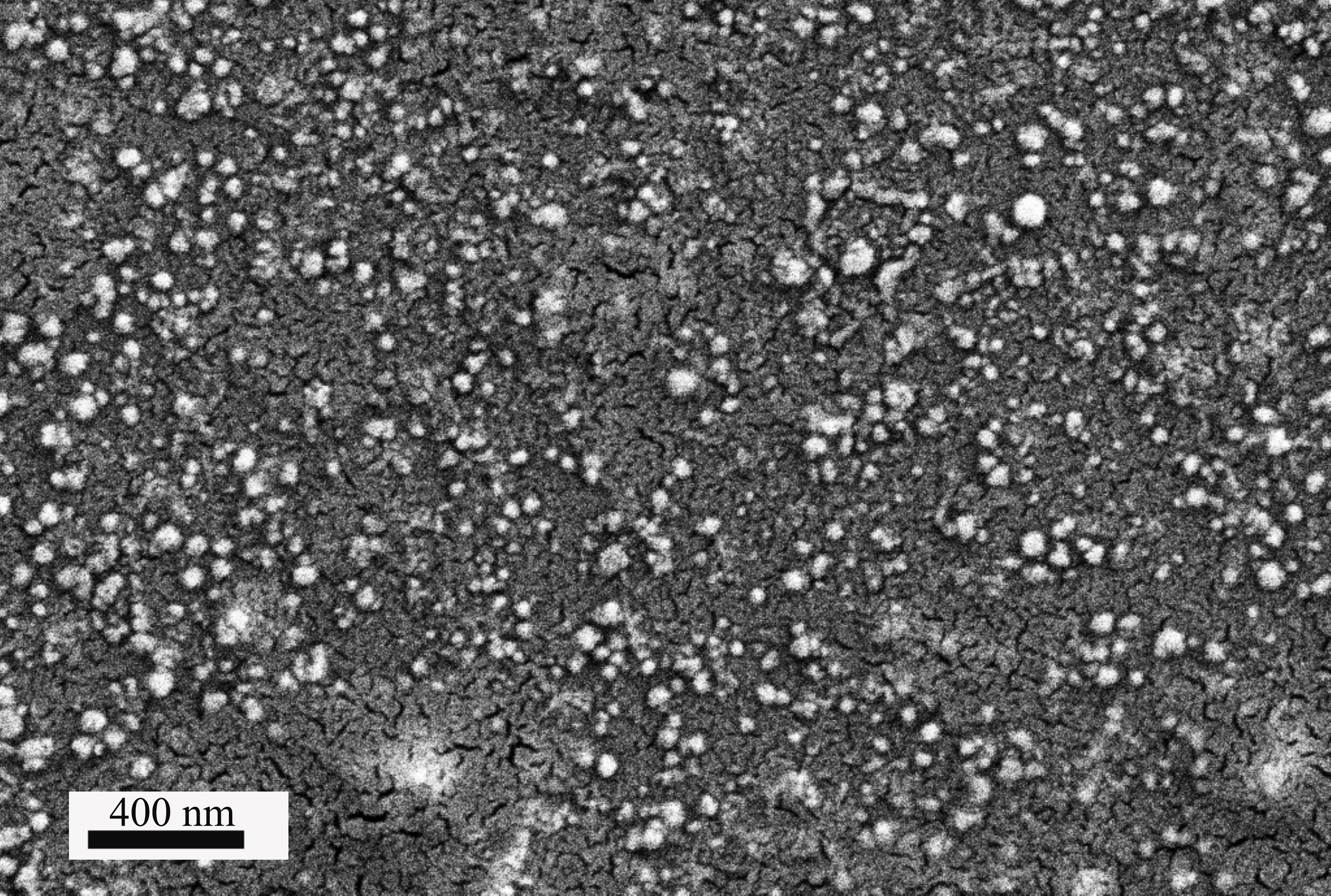}}%
\hspace{8pt}%
\subfigure[][]{%
\includegraphics[width=0.22\textwidth]{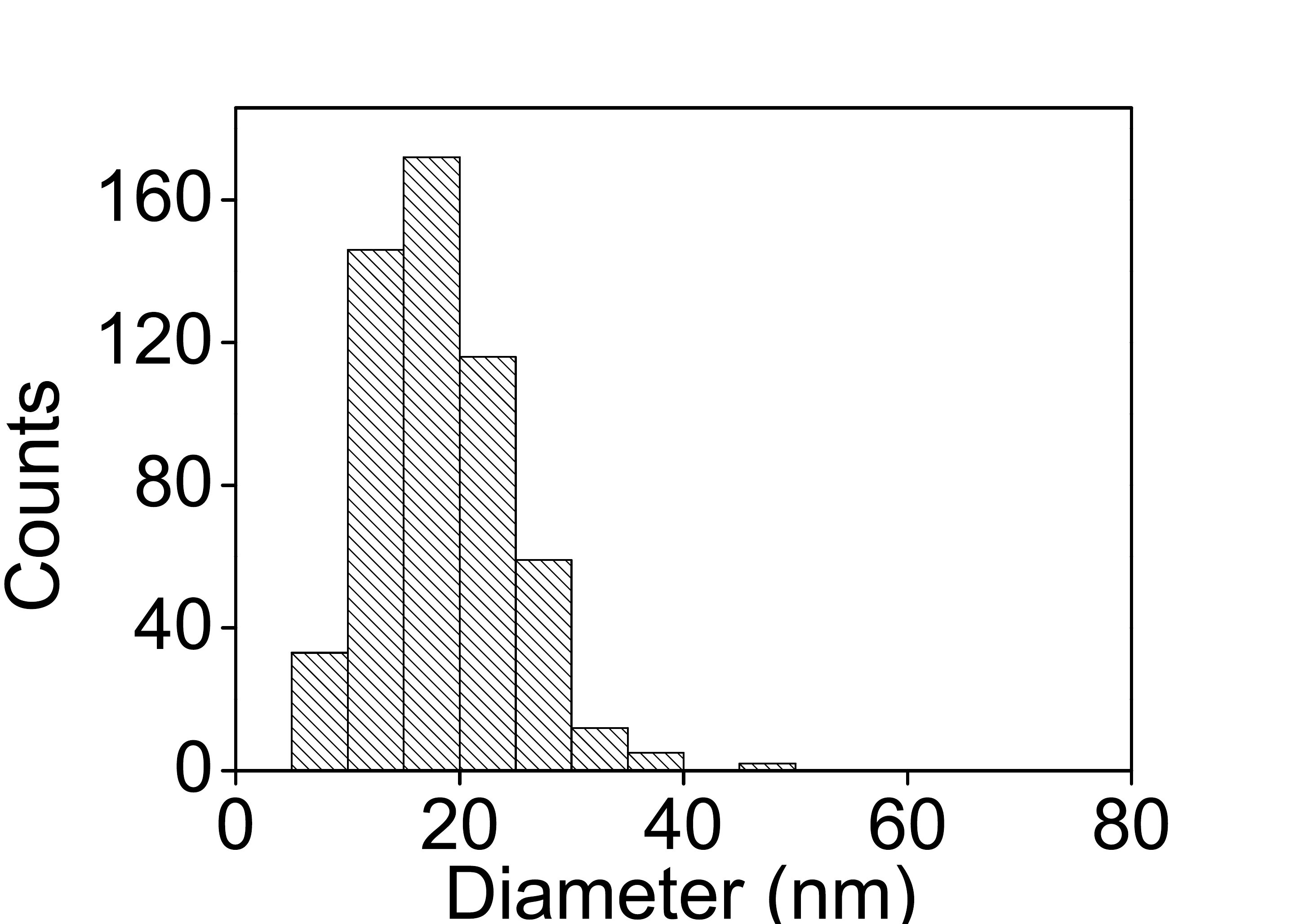}}%
\hspace{8pt}%
\subfigure[][]{%
\includegraphics[width=0.20\textwidth]{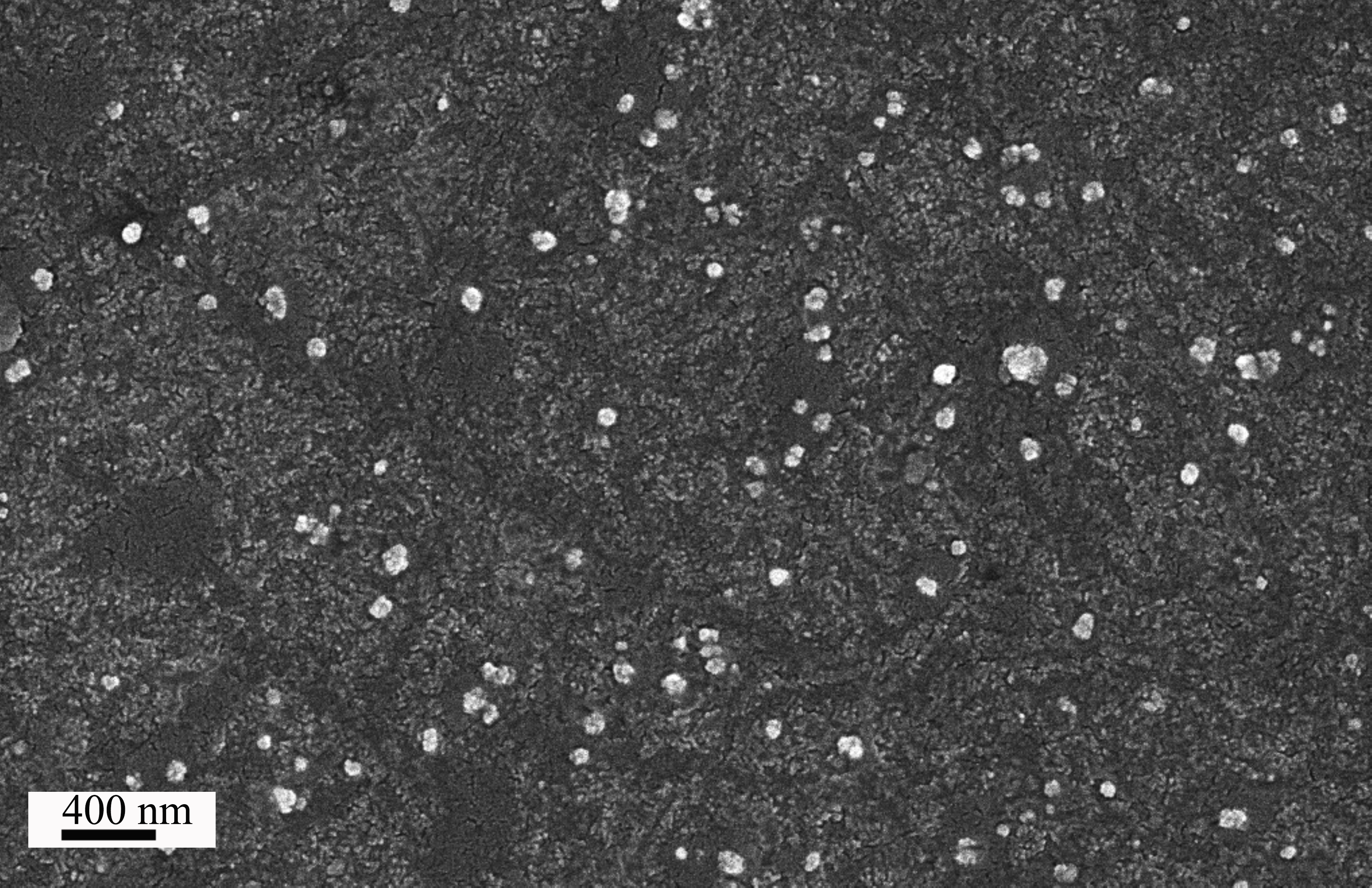}}%
\hspace{8pt}%
\subfigure[][]{%
\includegraphics[width=0.22\textwidth]{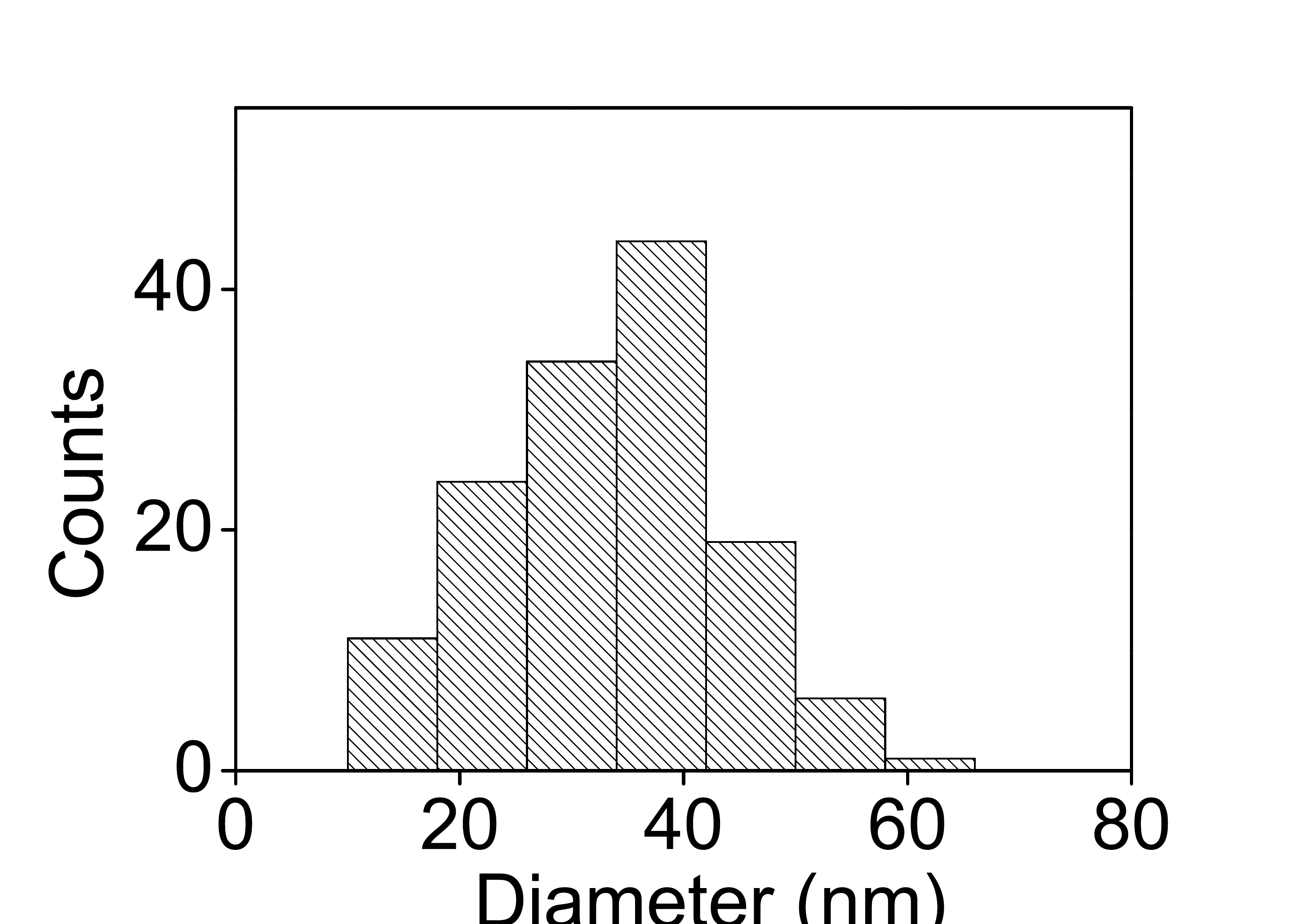}}%
\hspace{8pt}%
\subfigure[][]{%
\includegraphics[width=0.20\textwidth]{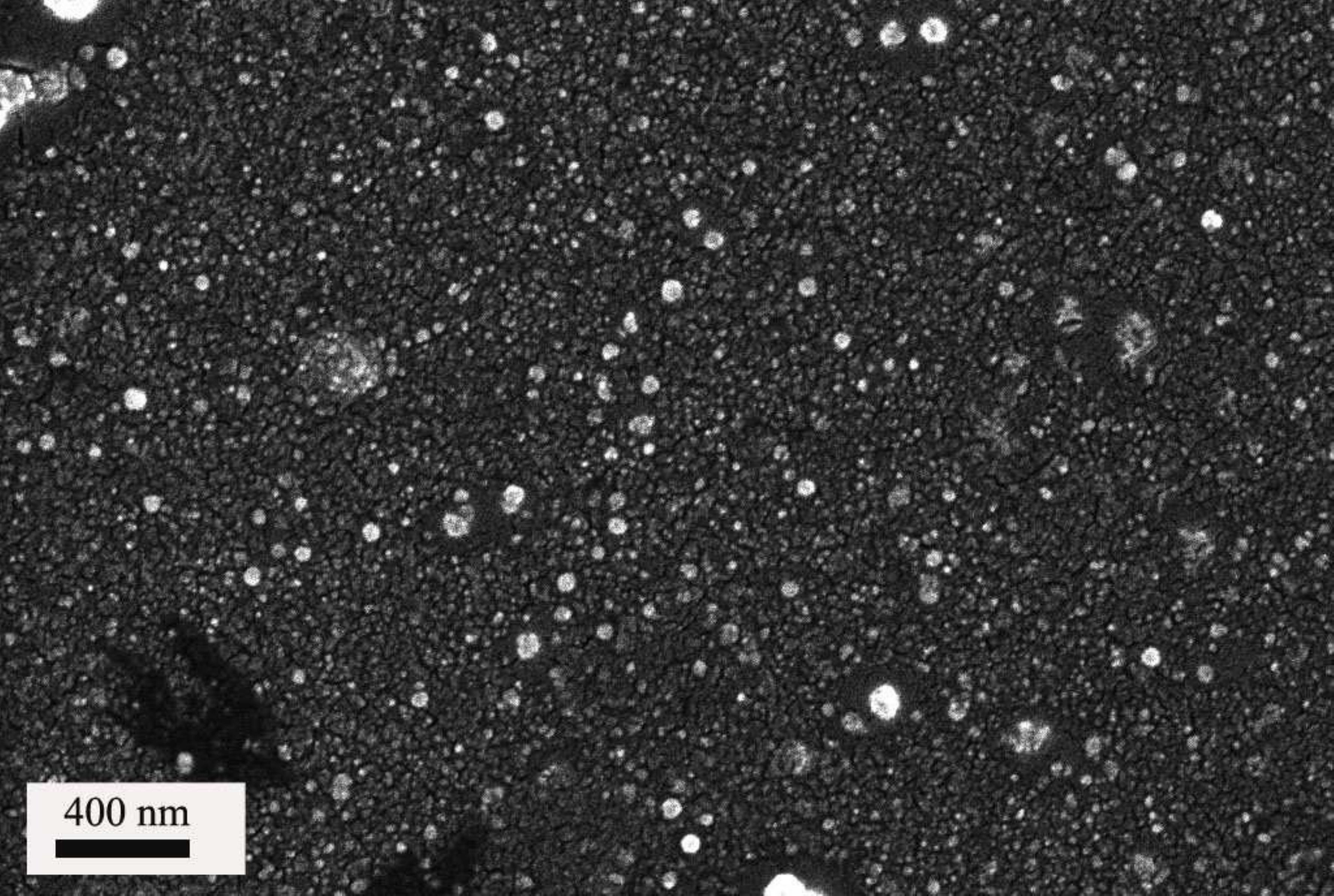}}%
\hspace{8pt}%
\subfigure[][]{%
\includegraphics[width=0.22\textwidth]{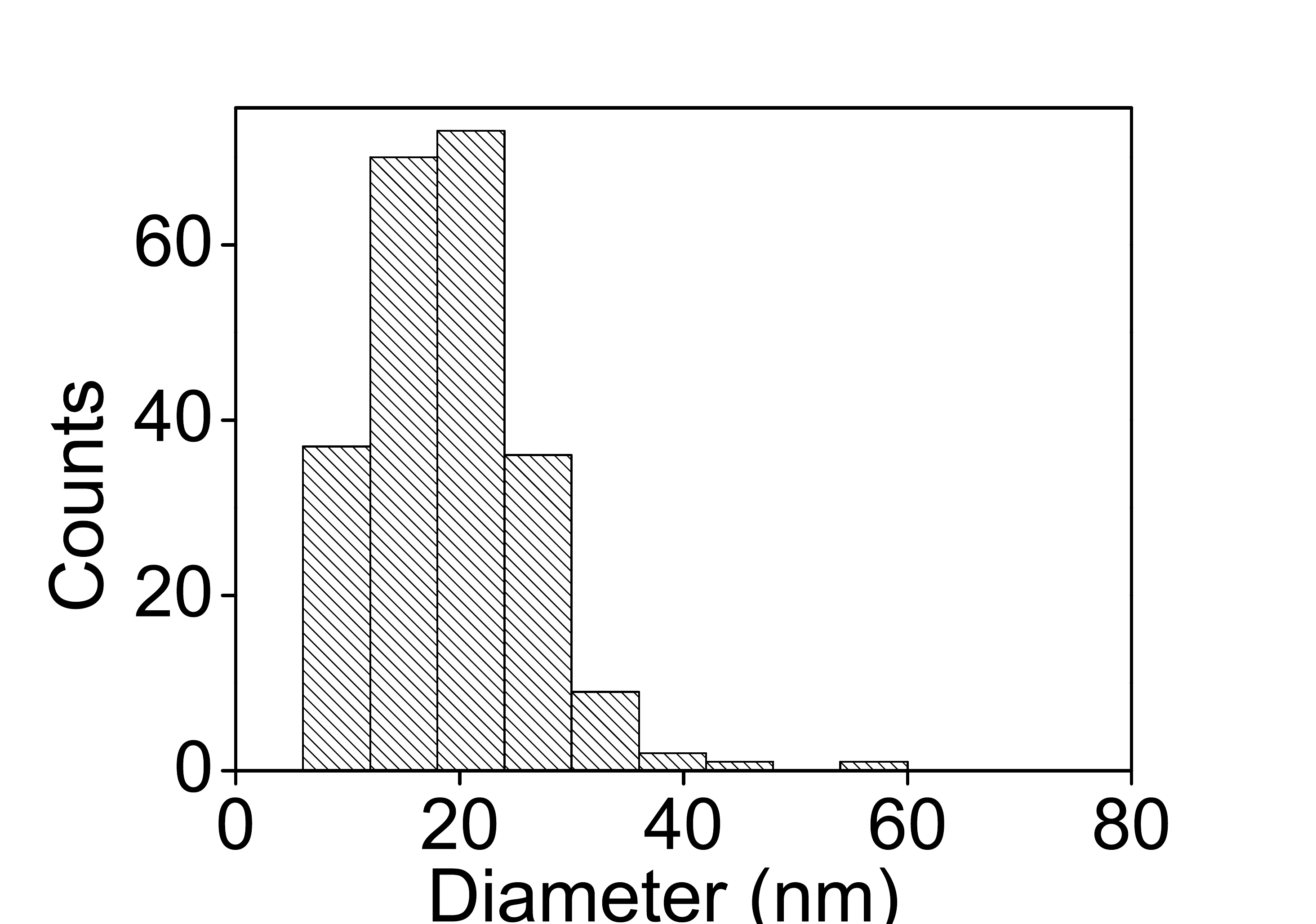}}%
\hspace{8pt}%

\caption[]{FESEM images of Ni$_{1-x}$V$_x$ samples with V compositions $x$ = 0.000 (a), 0.085 (c), 0.098 (e) and 0.11 (g). The corresponding particle size distributions are shown in (b), (d), (f) and (h), respectively.}%
\end{figure}

Figure 2 shows the FESEM images, along with the corresponding size distributions, of pure Ni and Ni$_{1-x}$V$_x$ alloy samples. As can be seen from the figure, the particle sizes range between 18$\pm$6 nm to 33$\pm$12 nm for the various samples.

\begin{figure}%
\centering
\subfigure[][]{%
\includegraphics[width=0.20\textwidth]{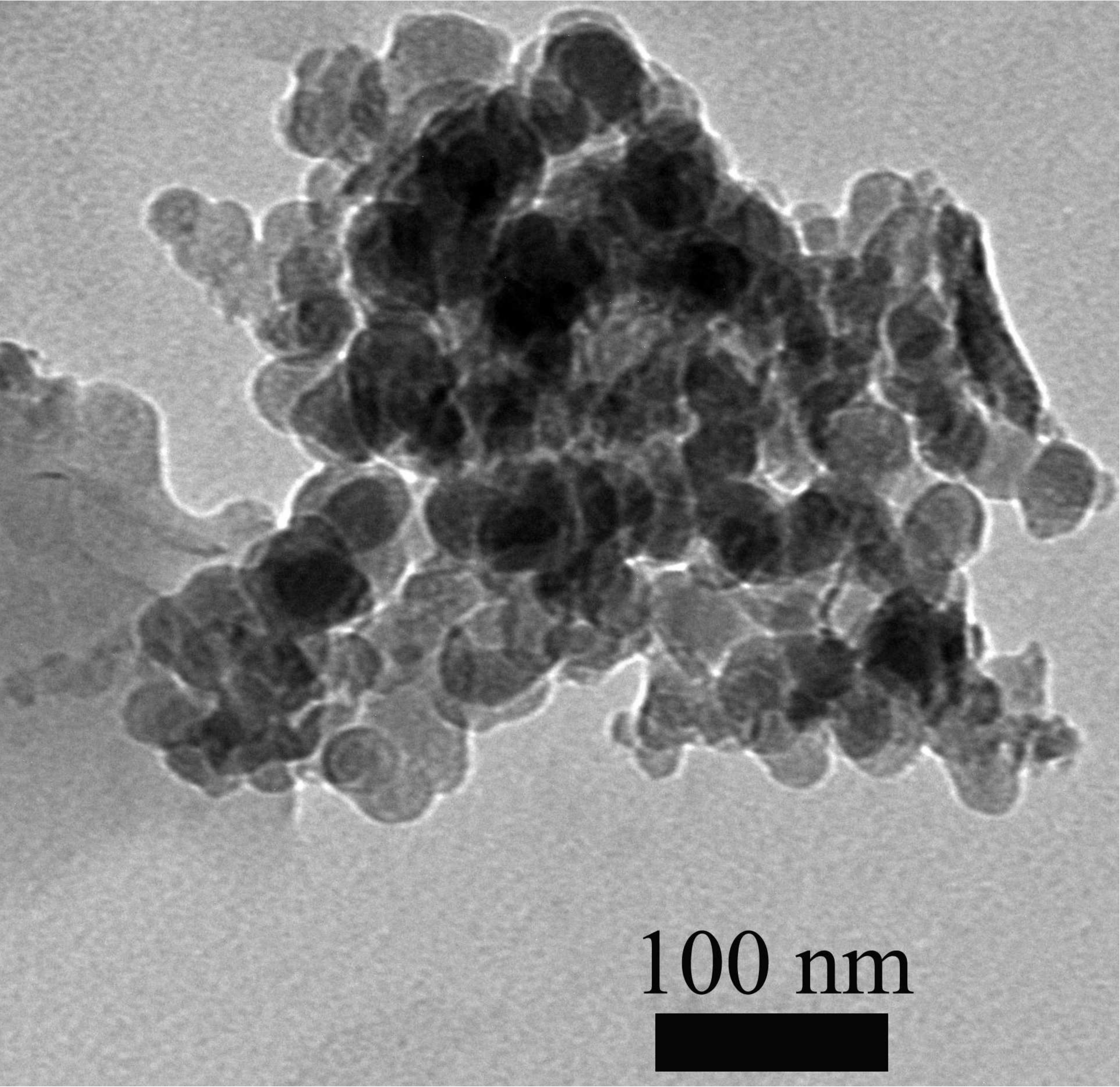}}%
\hspace{8pt}%
\subfigure[][]{%
\includegraphics[width=0.20\textwidth]{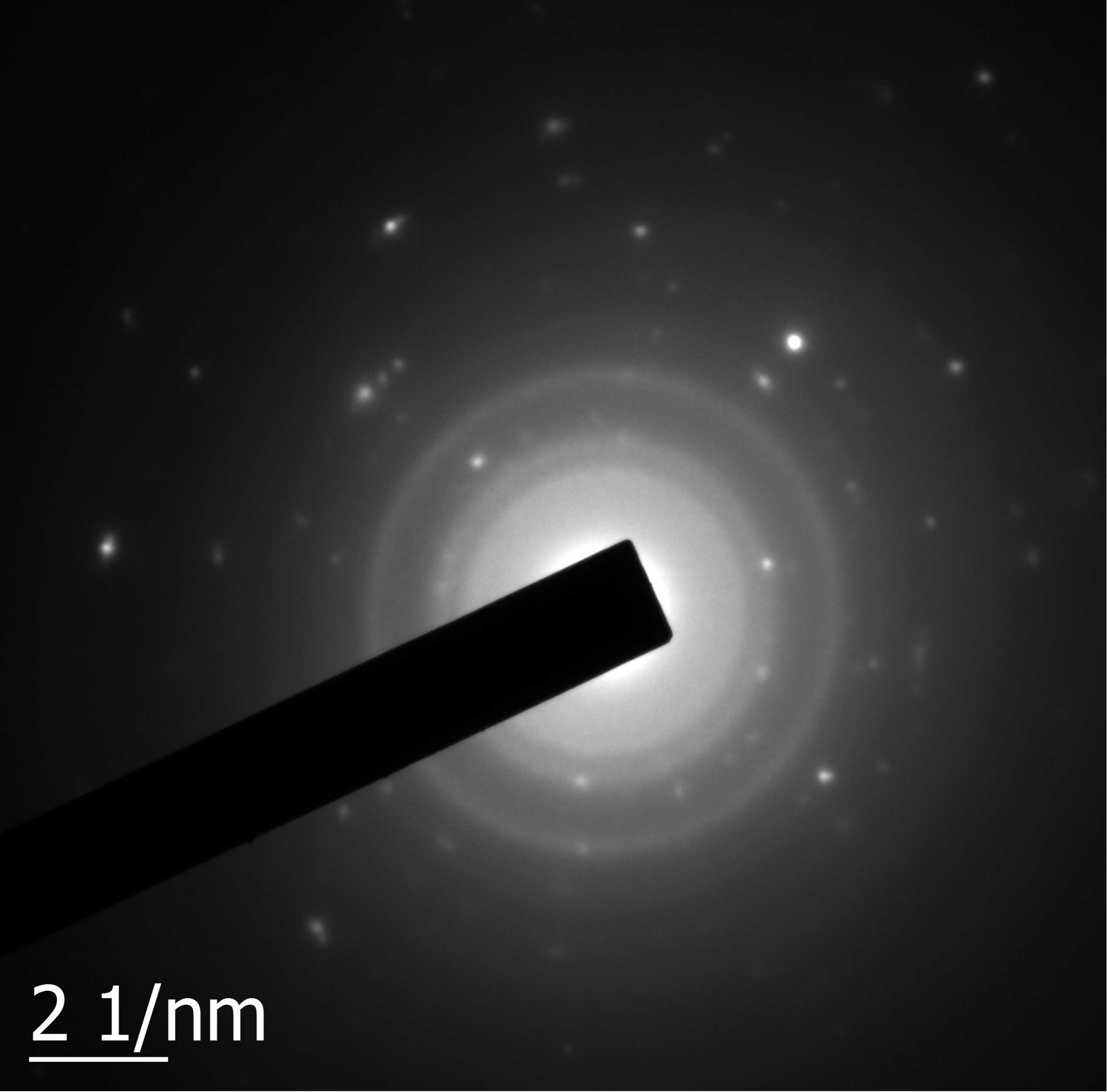}}%
\hspace{8pt}%
\subfigure[][]{%
\includegraphics[width=0.20\textwidth]{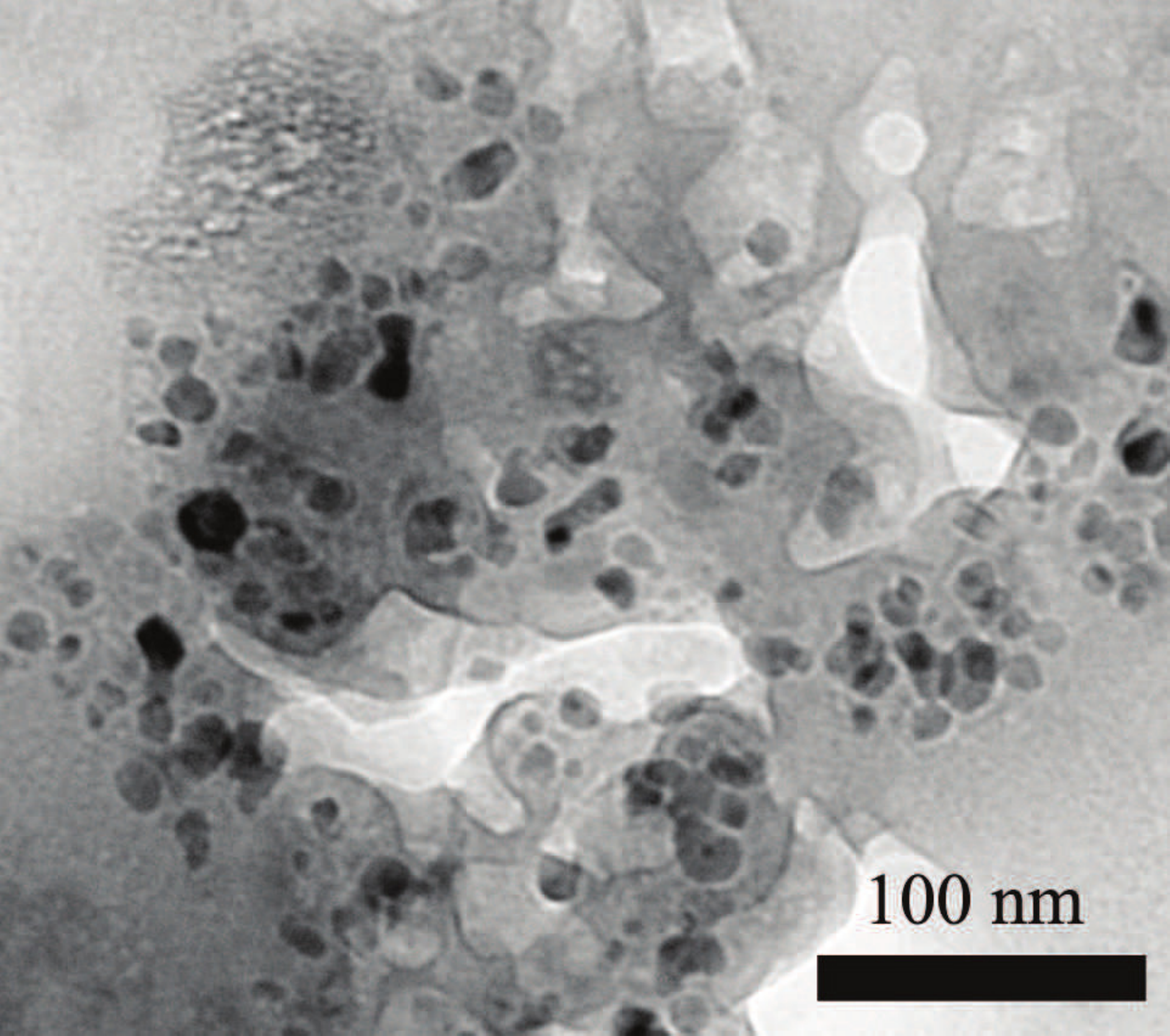}}%
\hspace{8pt}%
\subfigure[][]{%
\includegraphics[width=0.20\textwidth]{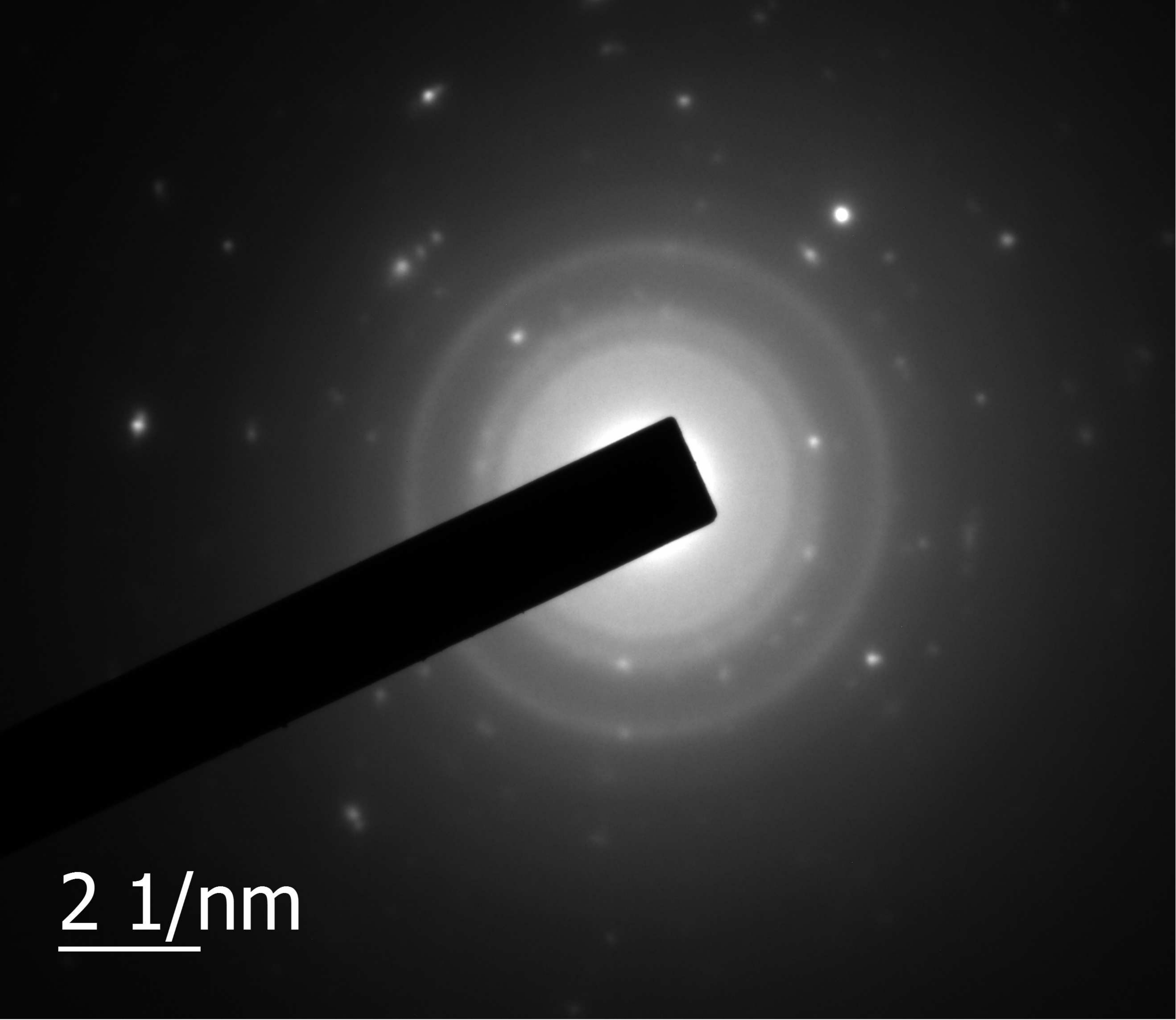}}%
\hspace{8pt}%
\subfigure[][]{%
\includegraphics[width=0.20\textwidth]{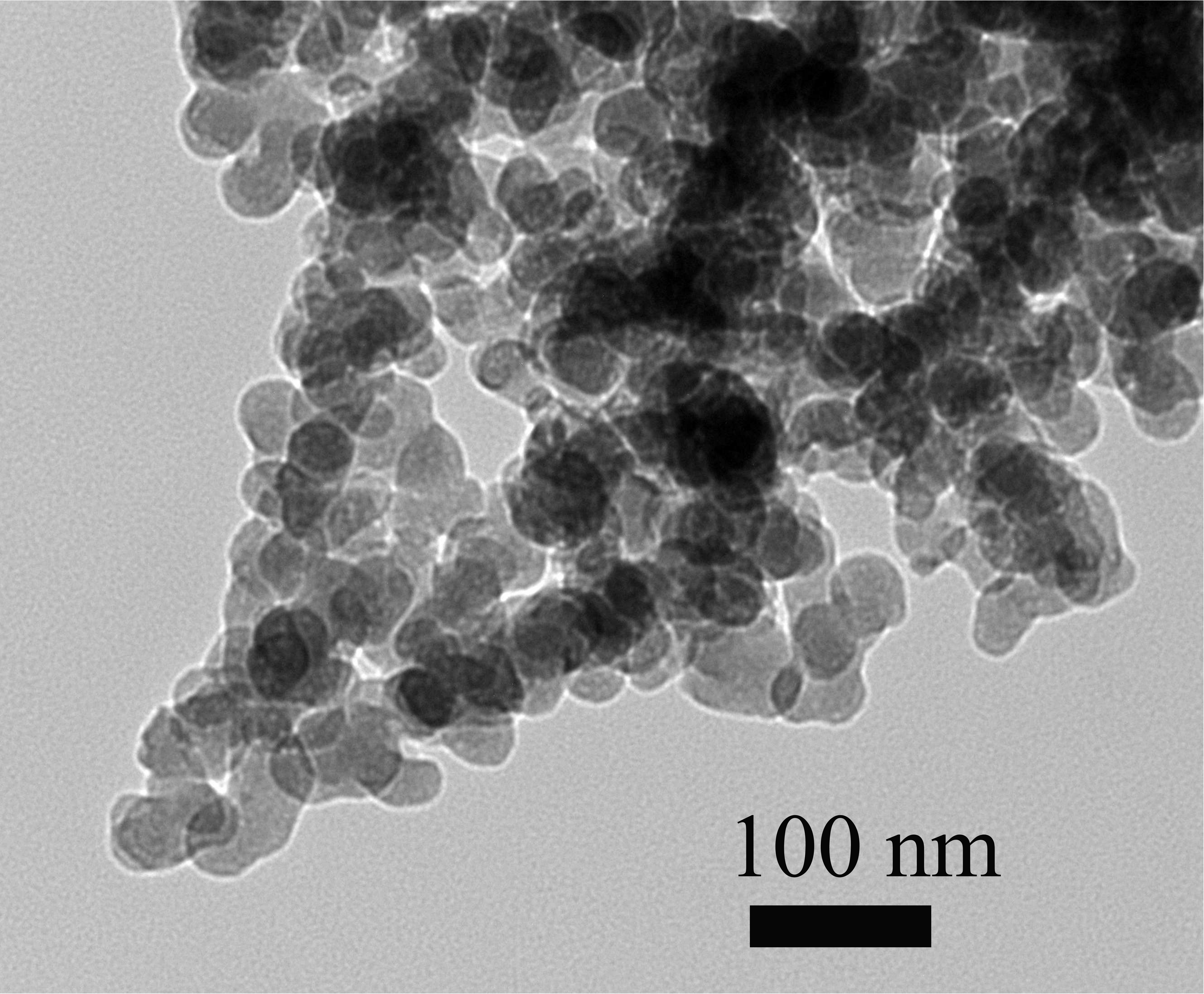}}%
\hspace{8pt}%
\subfigure[][]{%
\includegraphics[width=0.20\textwidth]{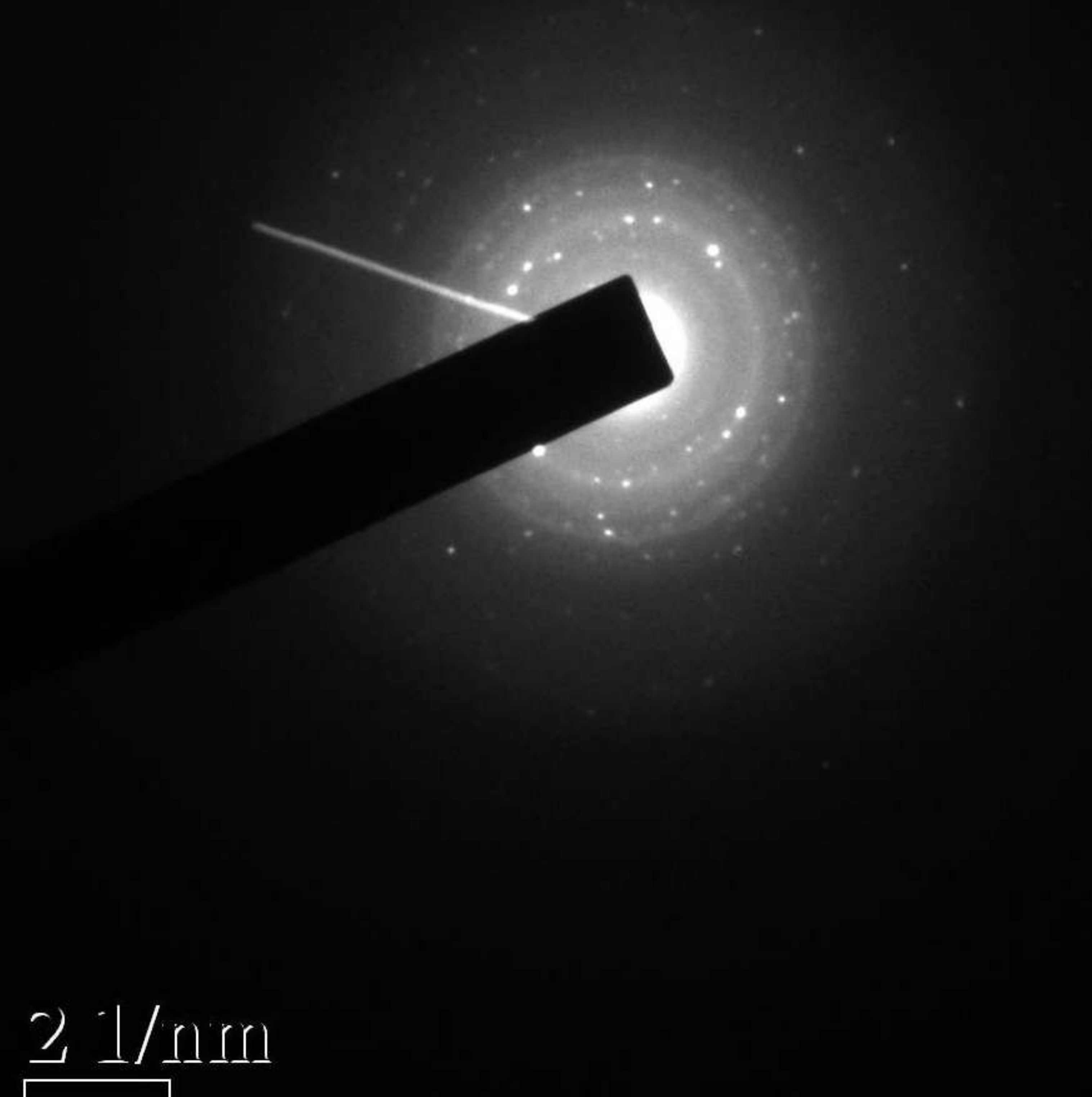}}%
\hspace{8pt}%

\caption[]{HRTEM images of Ni$_{1-x}$V$_x$ samples with V compositions $x$ = 0.000 (a), 0.098 (c) and 0.11 (e). The corresponding SAED patterns are shown in (b), (d) and (f), respectively.}%
\end{figure}

To analyze the particles further, HRTEM images and corresponding SAED patterns were taken for three representative samples with $x$ = 0 (pure Ni), 0.098 and 0.11. The images and patterns are shown in Fig. 3. The clearly visible nanoparticle agglomeration, which is not present in the corresponding FESEM images, indicates that the particles are magnetic.\cite{ABC} The occurrence of dots and concentric rings in each SAED pattern further elucidates that the particles are crystalline in nature.

\subsection{XRD}

\begin{figure}%
\centering
\subfigure[][]{%
\includegraphics[width=0.4 \textwidth]{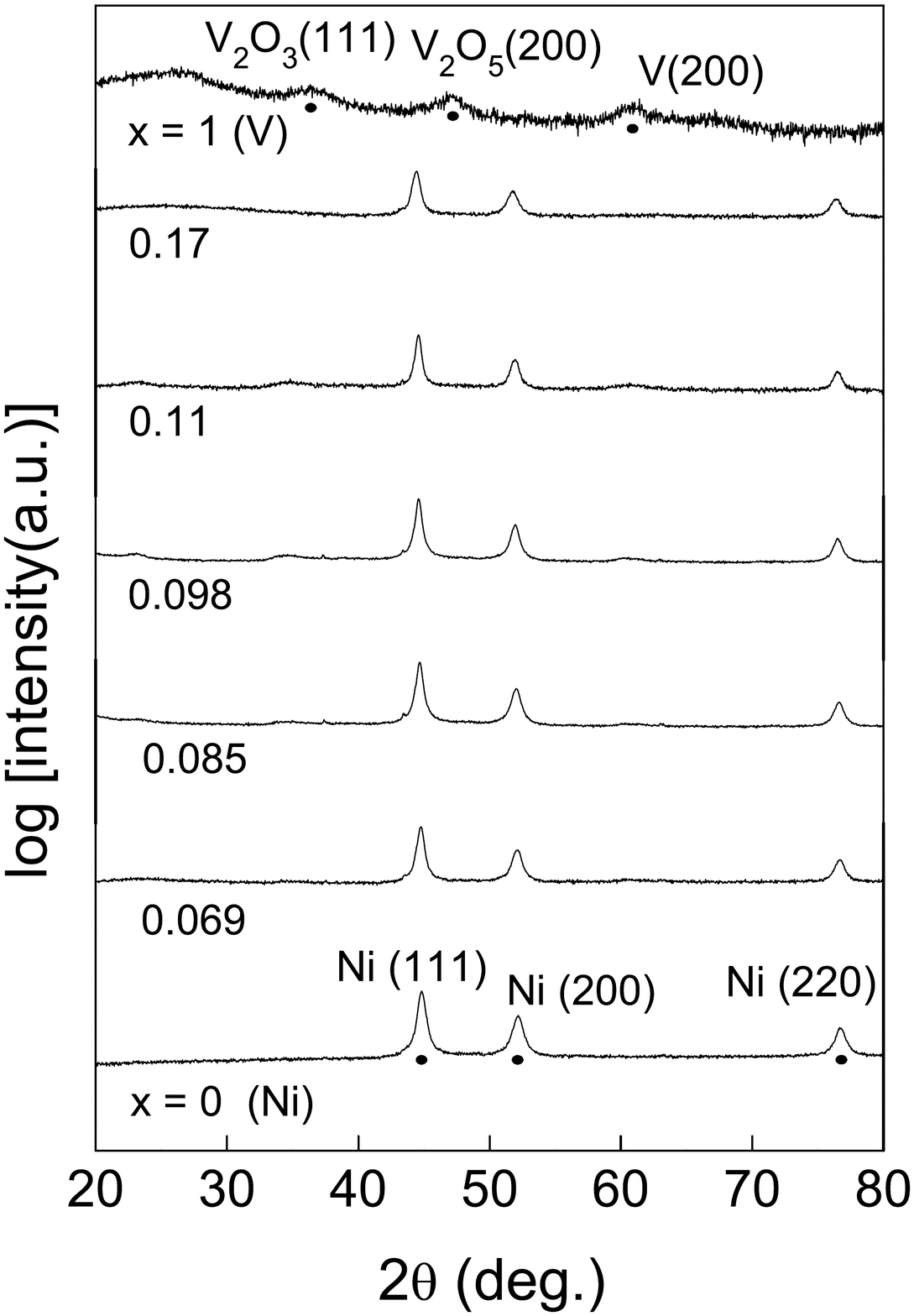}}%
\hspace{8pt}%
\subfigure[][]{%
\includegraphics[width=0.45\textwidth]{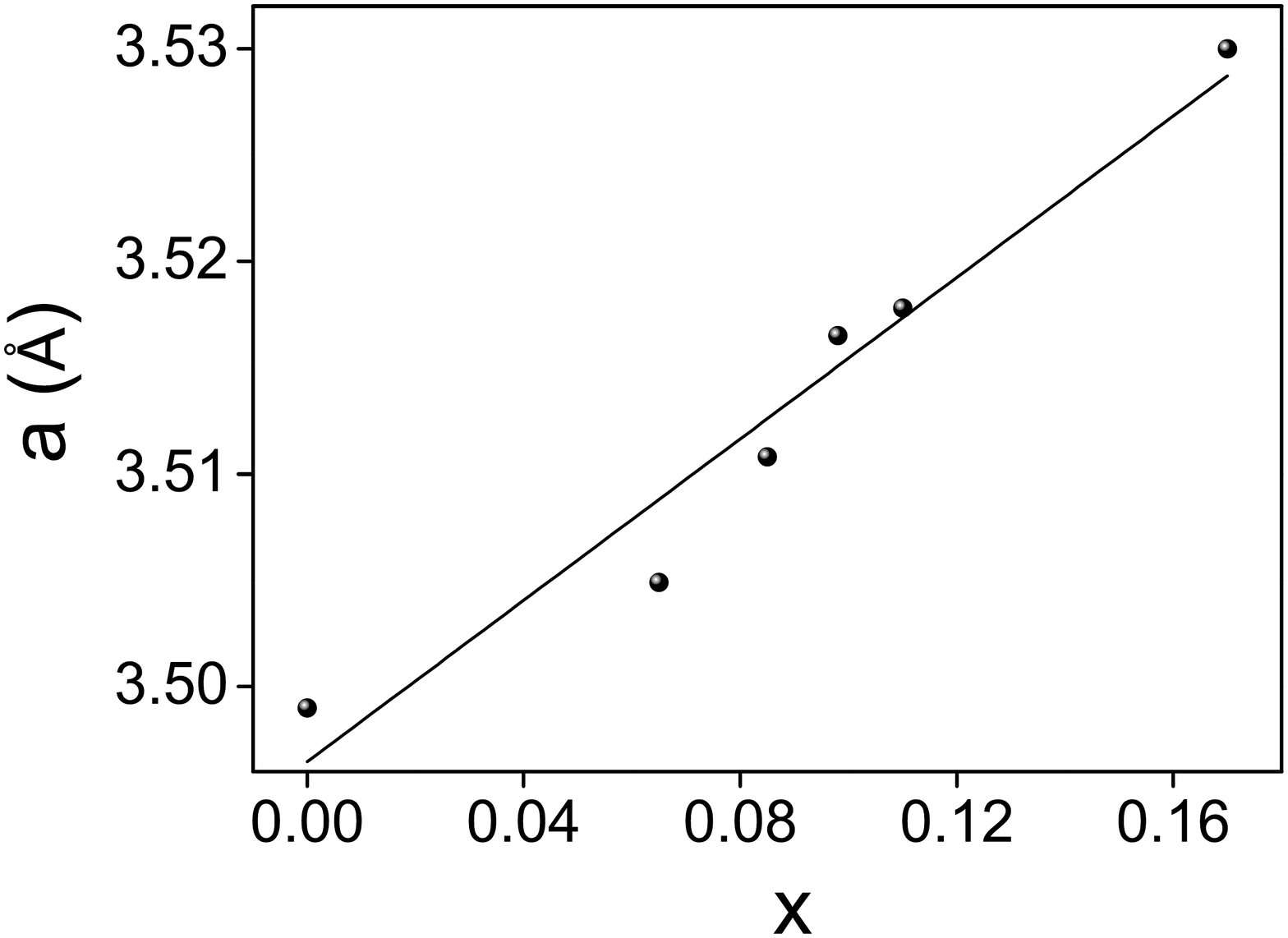}}%
\hspace{8pt}%

\caption[]{(a) XRD patterns of Ni$_{1-x}$V$_x$ samples. (b) Variation of lattice constant $a$ with $x$.}

\end{figure}

Figure 4(a) shows the XRD patterns of all the Ni$_{1-x}$V$_x$ (0 $\leq$ x $\leq$ 1) samples studied. The pattern for the pure Ni nanoparticles consists of three clean peaks at 44.50$^{\circ}$, 51.84$^{\circ}$ and 76.36$^{\circ}$. According to the Joint Committee on Powder Diffraction Standards (JCPDS) data, these peaks correspond to (111), (200) and (222) reflections of fcc Ni, but relatively displaced towards 2$\theta$ values higher than the corresponding bulk. This observation reveals that the Ni nanoparticles are pure in phase and are of a lattice constant (3.499 \AA) smaller than the corresponding bulk (3.524 \AA) value. Such a reduction of lattice constants in nanoparticles has earlier been predicted and demonstrated.\cite{Qi} The addition of V atoms to Ni up to $x$ = 0.17 does not alter the three-peak structure, except that these peaks progressively shift towards lower 2$\theta$ values. Further, essentially no additional peak(s) appear on V incorporation. This suggests that the alloy nanoparticles also are in the fcc phase and hence are in the form of Ni-V solid solutions. The alloy lattice constants $a(x)$, as determined from the (111) peak positions, are plotted as a function of $x$ in Fig. 4(b), and seem to vary linearly with $x$. A linear fit yields $a(x)$ = 3.4964 + 0.1898 $x$. Assuming a close-packed accommodation of the impurities, and hence that $a(x)$ is proportional to the weighted average of the atomic radii $r_{Ni}$ and $r_V$ of Ni and V, respectively, the ratio $r_{Ni}$/$r_V$ from the fitting parameters comes out to be 1.054. This value is close to 1.033, the covalent radius ratio for Ni and V, as available easily on the internet. This once again confirms the above inference that the Ni-V alloys are basically Ni-V solid solutions.

Notably, the solid solubility of V (17 $\%$) in the Ni nanoparticles is more than the solubility limit ($\leq$ 14 $\%$) for the bulk according to the Ni-V phase diagram.\cite{PhaseD} Solid solubilities in nanophase have earlier also been reported to be enhanced with respect to the bulk,\cite{Guo} and favour our results. The highly oxidized V nanoparticles, as evidenced by the dominant oxide peaks at 36.06$^{\circ}$ corresponding to V$_2$O$_3$ (111) and at 51.06$^{\circ}$ corresponding to V$_2$O$_5$ (200), were not studied further anyway, but helped confirm the absence of any oxide of vanadium in all other samples.

\subsection{XPS}

The survey XPS spectra of all the studied samples are shown in Fig. 5(a). The presence only of Ni and V peaks, apart from the adventitious C 1$s$ peak at 284.5 eV and one O (1$s$) peak at 531 eV, shown later not to participate in any oxide formation except in the pure V nanoparticle case, corroborates the XRD results on the high purity of the pure Ni and alloy nanoparticles. Further, the high-resolution XPS spectra in Ni and V binding energy (BE) regions are shown in Figs. 5(b) and 5(c), respectively. The non-deconvolutable single spin-orbit split peaks in each case confirm the absence of any oxide in the samples. Further, the peaks shift towards higher BE with respect to the pure Ni nanoparticle values on increasing $x$ in both the sets. The peak shifts ($\Delta_{\rm BE}$) are then plotted as a function of $x$ in Fig. 5(d). Both the Ni and V peaks can be seen to vary monotonically and essentially concurrently with $x$, confirming the Ni-V alloy formation with different V concentrations, as also reported earlier.\cite{Steiner1981, swain15}

\begin{figure}%
\centering
\subfigure[][]{%
\includegraphics[width=0.45\textwidth]{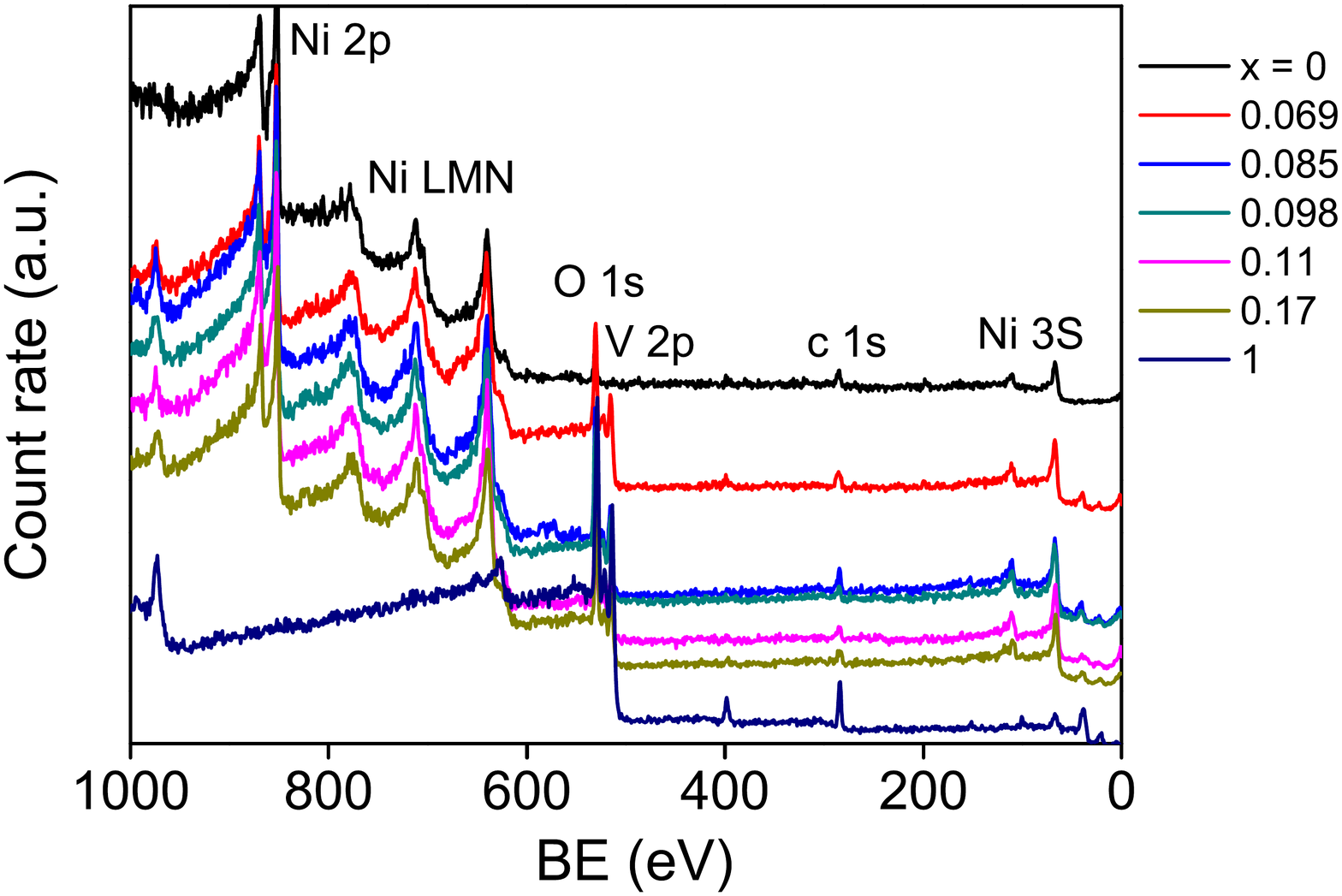}}%
\hspace{8pt}%
\subfigure[][]{%
\includegraphics[width=0.22\textwidth]{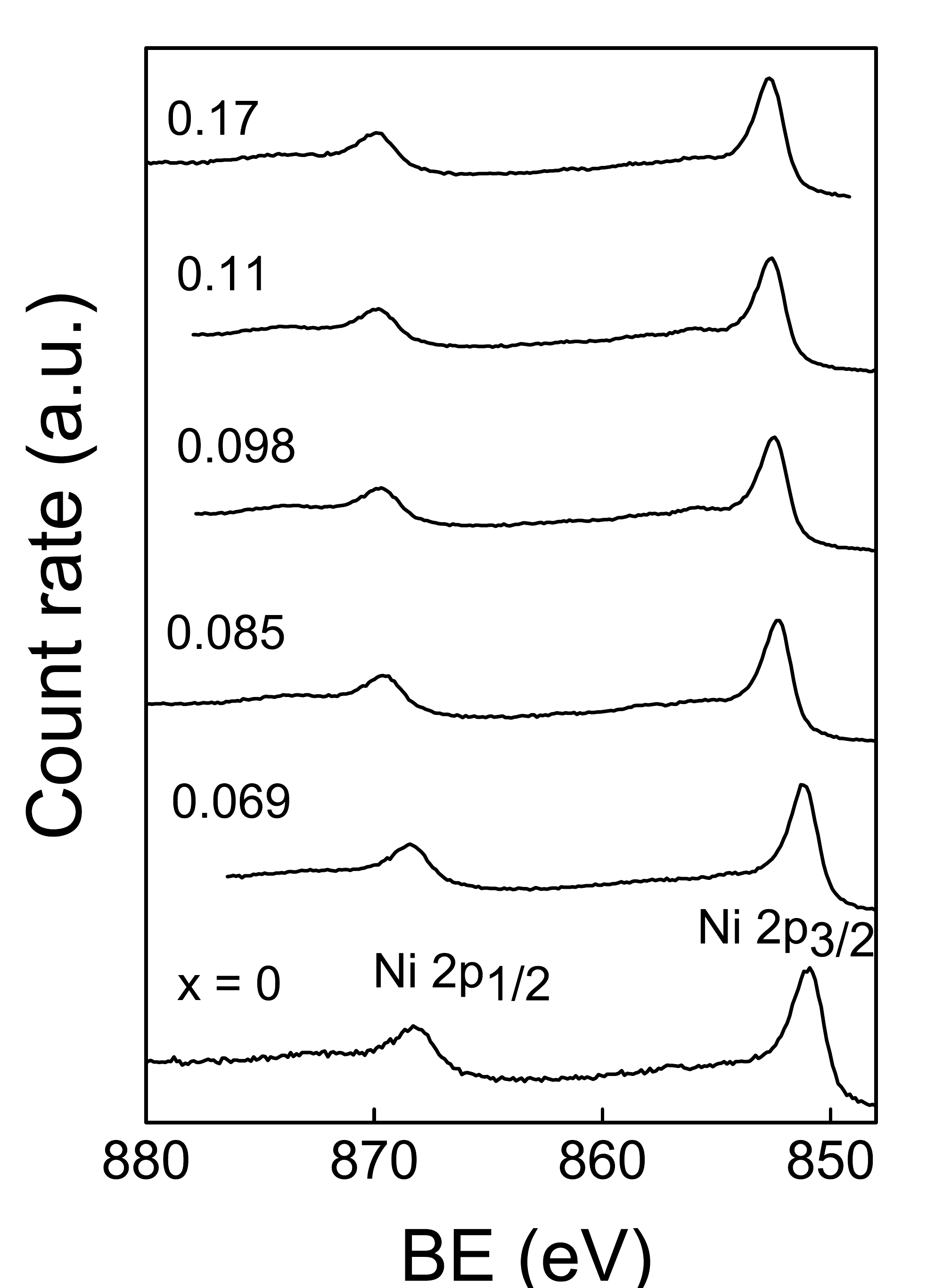}}%
\hspace{8pt}%
\subfigure[][]{%
\includegraphics[width=0.22\textwidth]{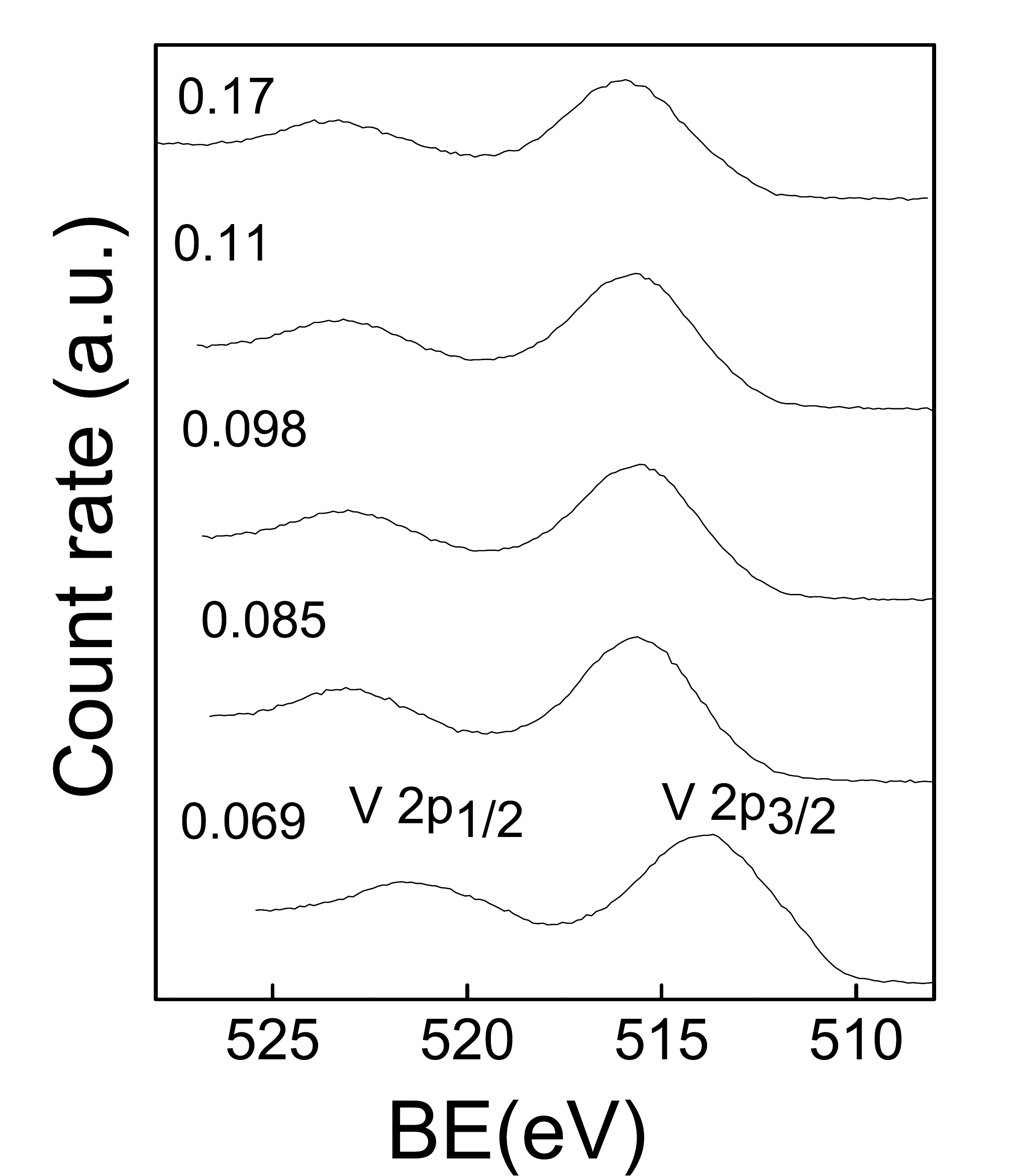}}%
\hspace{8pt}%
\subfigure[][]{%
\includegraphics[width=0.35\textwidth]{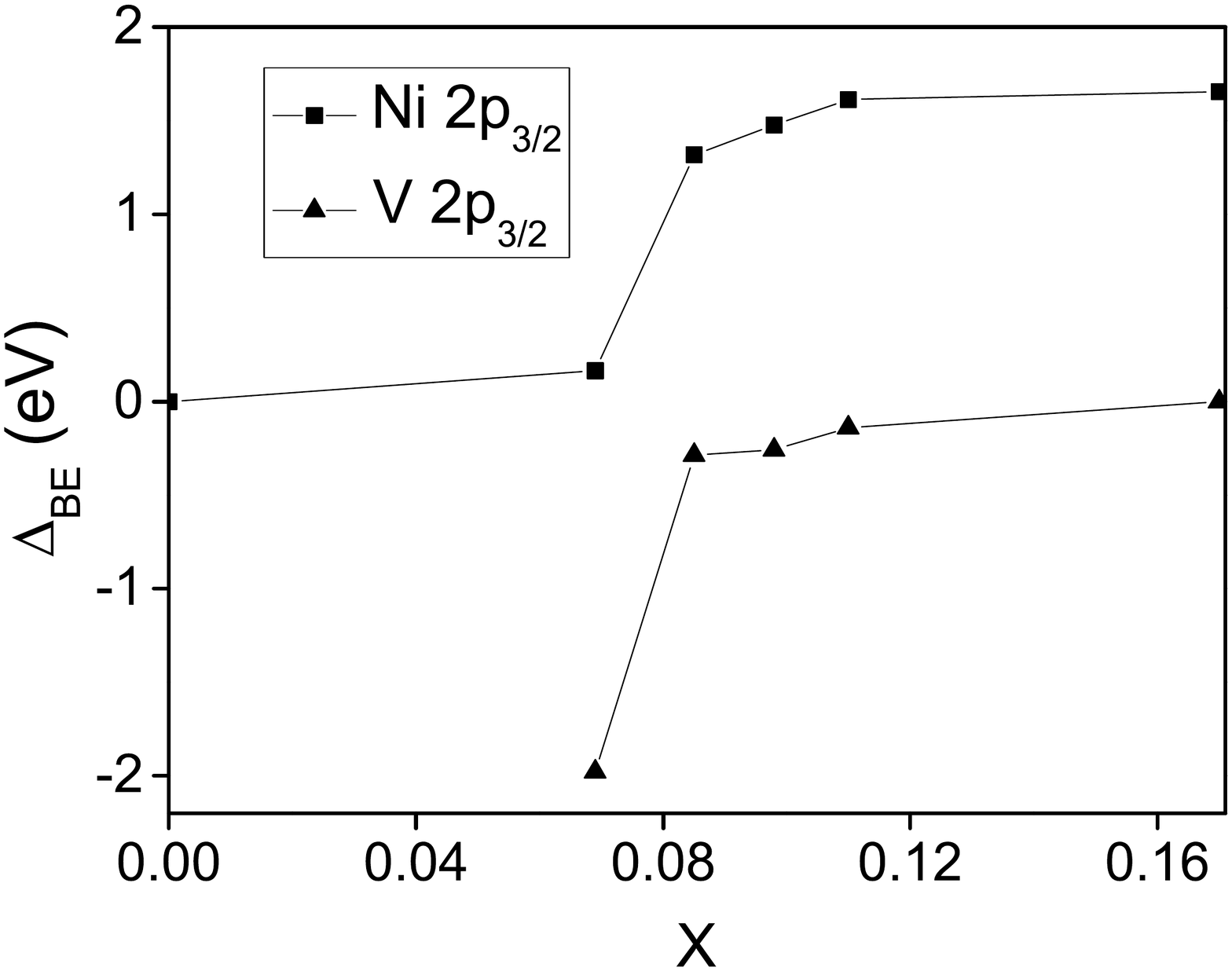}}%
\hspace{8pt}%

\caption[]{(a) Survey XPS spectra of Ni$_{1-x}$V$_x$ samples. (b) High-resolution XPS spectra in the Ni 2p region. (c) High-resolution XPS spectra in the V 2p region. (d) Variation of Ni and V peak positions with $x$.}
\end{figure}

\subsection{Resistivity}

The residual resistivity ratio (RRR), defined as RRR = [$\rho$ (T) - $\rho$ (15 K)]/$\rho$ (15 K), has been plotted as a function of temperature for four representative samples in Fig. 6; $\rho$ (T) here is the resistivity at temperature T. From the figure, it can be seen that apart from the monotonic increase of resistivities confirming the metallic nature of the alloy nanoparticles and ruling out their oxidation, each curve shows a peak in 40 K - 60 K temperature range. This peak is indicative of the presence of a small nanoparticle volume with uncorrelated PM-like spins, which start getting gradually aligned in the field direction at this temperature and on lowering the temperature further, in line with the observations reported earlier.\cite{Diep2011, Magnin2011}

\begin{figure}%
\centering
\includegraphics[width=0.45\textwidth]{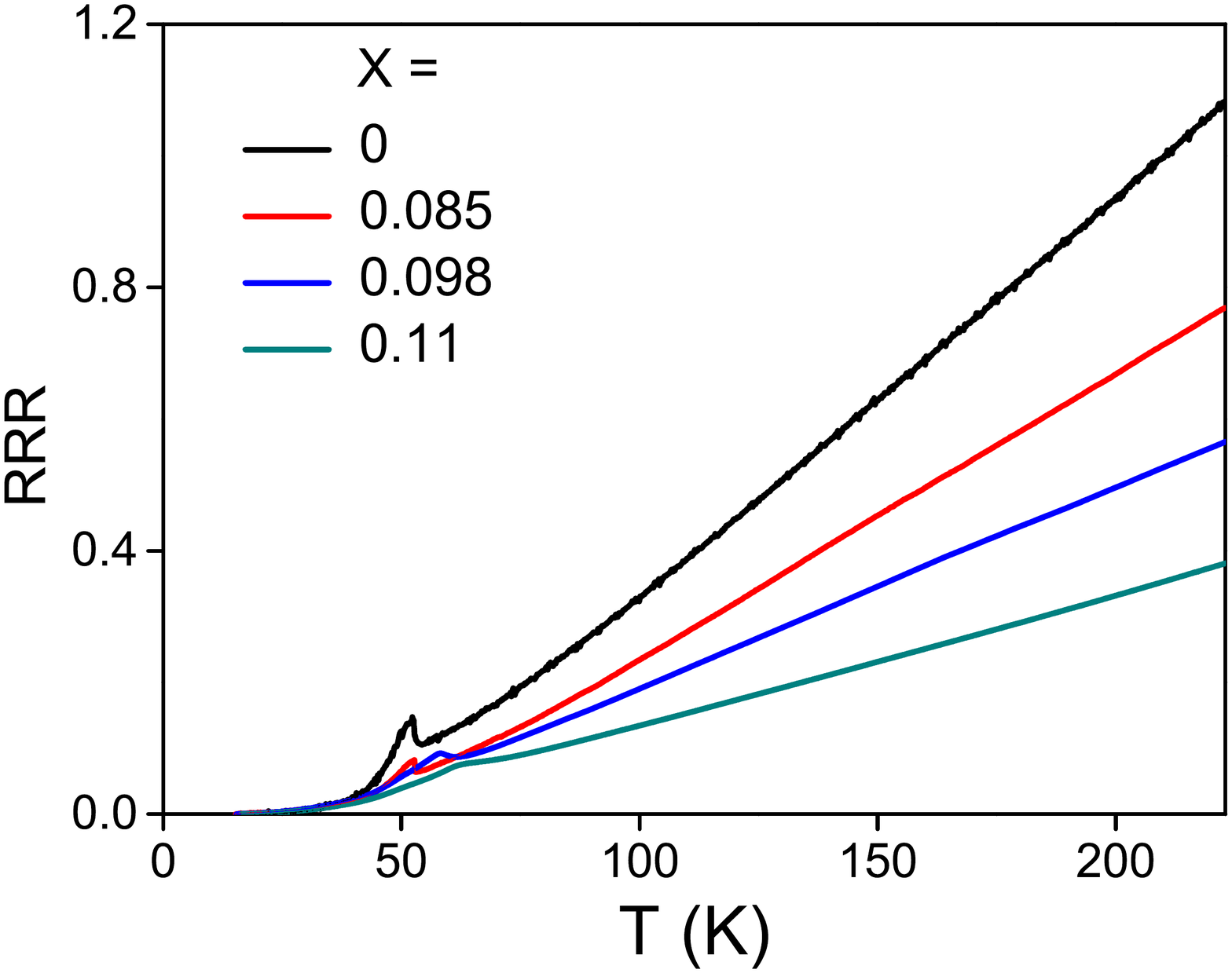}%
\hspace{8pt}%

\caption[]{Temperature dependence of the residual resistivity ratio in the temperature range 15K - 230 K for $x$ = 0.000, 0.085,0.098, and 0.11.}%
\end{figure}

\subsection{Magnetization}

The zero-field cooled (ZFC) and field cooled (FC) dc magnetization versus temperature curves in the range 5 K $\leq$ T $\leq$ 300 K are plotted in Fig. 7 for all the samples. The curves are suitably shifted to coincide roughly at the same point at 300 K. The ZFC-FC splitting in all the cases is characteristic of SPM nature of nanoparticles,\cite{castrillon} and indicates that the nanoparticles with V concentration as high as 17 $\%$ are all magnetic. This is in agreement with the HRTEM observation of particle agglomeration as pointed out above. On the other hand, the fact that the Ni$_{0.83}$V$_{0.17}$ nanoparticles are also magnetic, is in discordance with the PM bulk behaviour of the alloy of this composition.\cite{Schroeder10} This discordance, however, is acceptable since it is well known that in many cases PM materials transform to magnetic phases in nanodimensions.

Despite the ZFC peak in every case being broad because of the wide particle size distribution\cite{Shim} as found in FESEM images, the blocking temperature T$_{\rm B}$ can still be determined by estimating the inflection point of the dM/dT versus T curve (see Fig. 7). Below T$_{\rm B}$, the nanoparticles are in a blocked FM state, which is also signalled by a kind of saturation of FC magnetization in this region.\cite{Lueker} T$_{\rm B}$ can be plotted as a function of $x$ to separate the SPM and blocked FM phases above and below it, respectively. A striking feature of both the ZFC and FC curves for each composition is the occurrence of a PM-like increase in magnetization below an $x$-dependent temperature T$_{\rm P}$ (see Fig. 7). This temperature can be estimated in the following manner: First, the initial part of the FC curve below T$_{\rm B}$ is extrapolated to the lowest measured temperature, and then the extrapolated curve is subtracted from the original FC curve. The difference would start rising up from zero at T$_{\rm P}$ when going down in temperature, as shown schematically in Fig. 8(a). Incidentally, the T$_{\rm P}$ estimated in this manner is quite close to the peak position in the corresponding resistivity curve. Figure 8(b) displays a relation between the two temperatures. Certainly, this temperature rise below T$_{\rm P}$ is indicative of the emergence of a PM-like phase inside the blocked FM nanoparticles. It is this phenomenon which had resulted in the occurrence of the peak in the corresponding resistivity curve (Fig. 6). For an estimation, the T$_{\rm P}$ and the corresponding resistivity peak position can be averaged to get a modified T$_{\rm P}$, which can then be plotted as a function of $x$ to draw a boundary in the phase diagram below which the PM-like phase coexists with the blocked FM phase. The phase diagram estimated this way is drawn in Fig. 9. This coexistence of the two phases at lower temperatures is also supported by the presence of a small hysteresis loop along with an unsaturated magnetization in the M-H curve of each sample at 2 K, as shown in Fig. 10.

Although the inference above on the coexistence of the blocked FM and PM-like phases below T$_{\rm P}$ is enough as an interpretation to the limited amount of data presented in this work, it would perhaps still be inequitable to ignore examining the shapes of the M-T curves below T$_{\rm P}$ further to some extent. For this purpose, it would be sufficient to investigate just the FC magnetization curves. As can be seen from Fig. 7, there are two kinds of patterns of the FC magnetization below T$_{\rm P}$: it either saturates below an even lower temperature as in the case for $x$ =0, or keeps on increasing down to the lowest measured temperature, as is observable for all other compositions.

Let us start from $x$ = 0 (pure Ni) case. The second near-saturation of magnetization below $\sim$ 10 K suggests that a small volume in each nanoparticle is still FM but with a weak exchange interaction ($\mathsf{J}$). A Curie-Weiss fit of the low-temperature magnetization results in the Curie temperature (T$\rm{_C}$) and Curie constant (C) values of 15 K and 0.14 K, respectively. Further, the derived Weiss field constant $\lambda$ = 115, which is a measure of $\mathsf{J}$,\cite{Blundell} is much lower than its value $\sim$ 5000 for FM bulk Fe, and substantiates the argument that the exchange interaction in this volume is indeed weak. This observation is in line with a report by Qin {\it et al.},\cite{Qin2014} according to which a superparamagnetic nanoparticle may have a core region with strong $\mathsf{J}$ surrounded by spins with weaker $\mathsf{J}$ at the surface. With decreasing temperature, the fluctuations of the surface spins slow down and a short range correlation grows between them, giving rise to a sharp increase in magnetization, which saturates below T$\rm{_C}$ when all the surface spins are aligned along the field direction. A schematic of the magnetic structure of pure Ni nanoparticles is shown in Fig. 9.

\begin{figure}%
\centering
\includegraphics[width=0.5\textwidth]{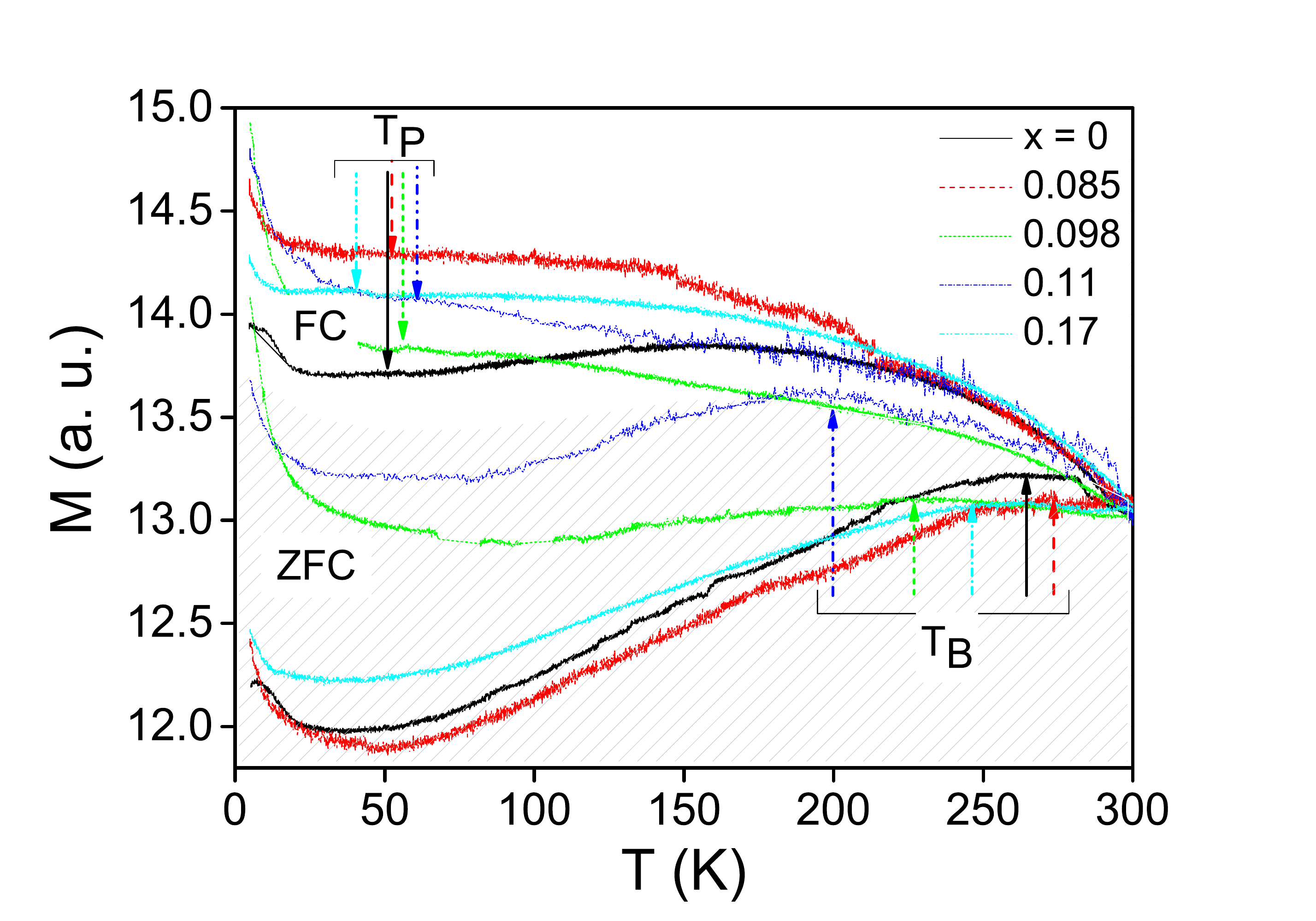}%

\caption[]{FC and ZFC magnetizations versus temperature at 500 Oe field for $x$ = 0.000, 0.085, 0.098, 0.11 and 0.17. The region containing essentially the ZFC curves is hatched for visual separation between FC and ZFC curves.}%
\end{figure}

\begin{figure}%
\centering
\subfigure[][]{%
\includegraphics[width=0.4\textwidth]{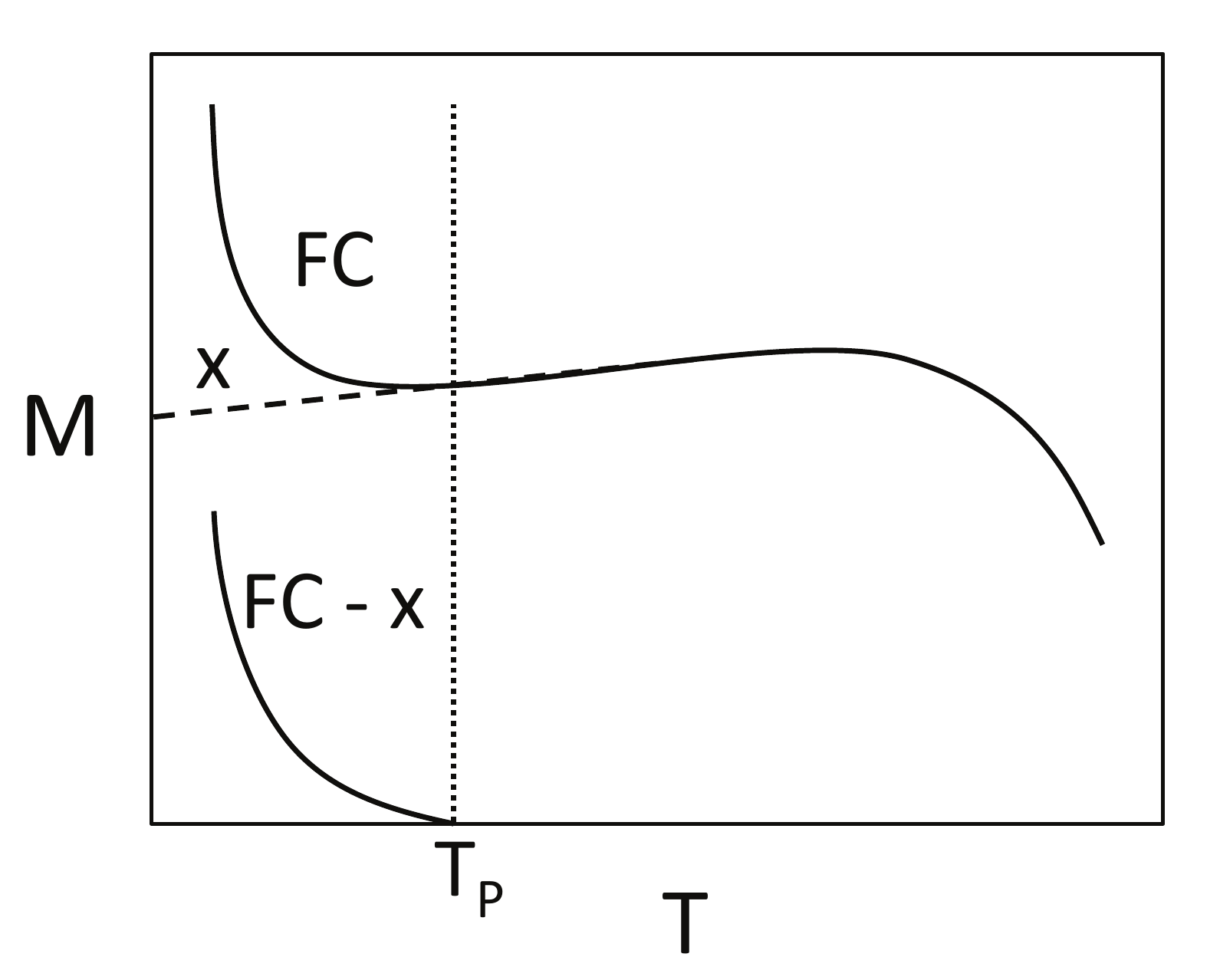}}%
\hspace{8pt}%
\subfigure[][]{%
\includegraphics[width=0.5\textwidth]{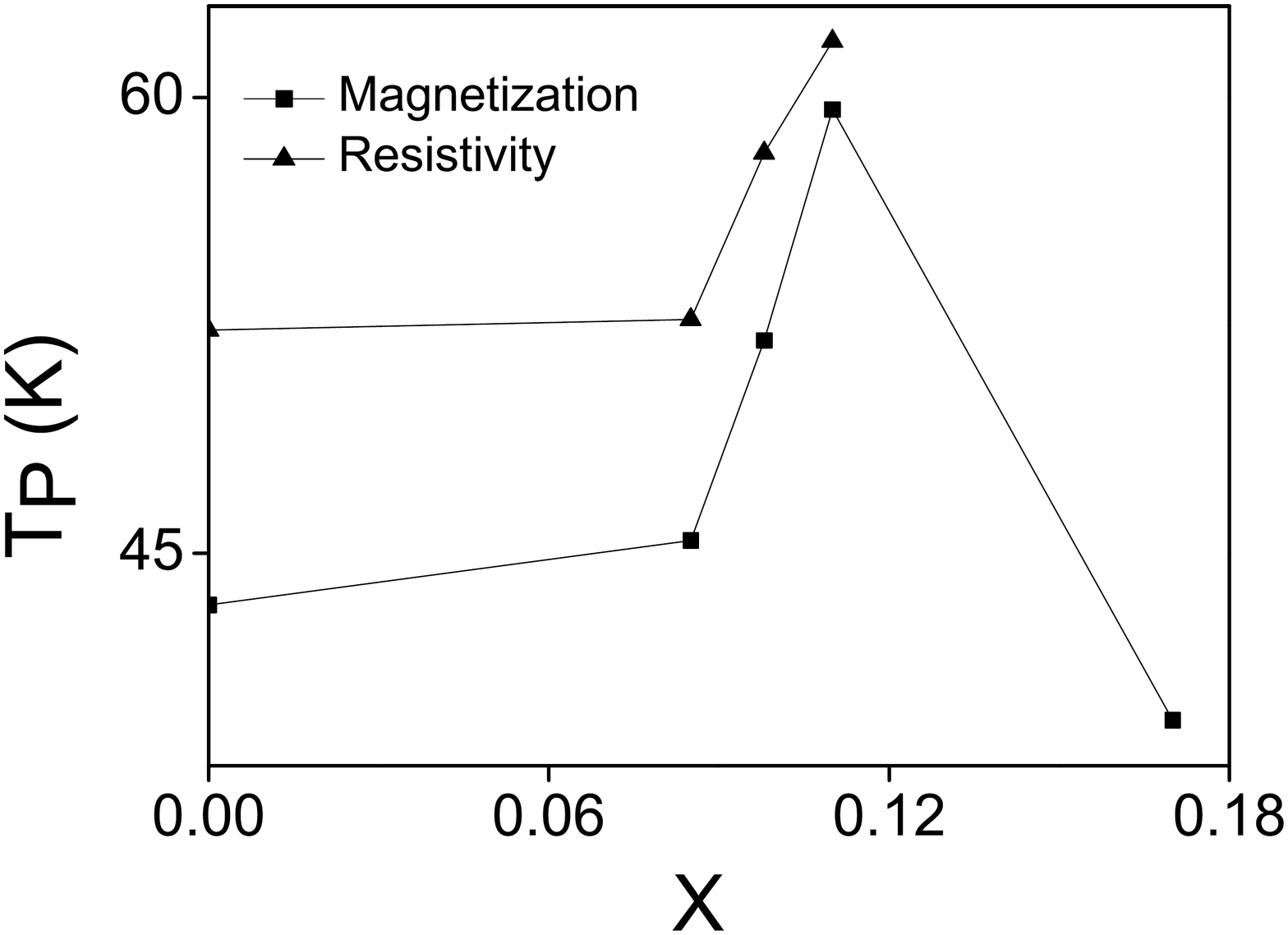}}%

\caption[]{(a) Schematic of a typical FC magnetization curve (upper curve), the interpolation (x) as described in the text, and the difference (FC-x) with an identification of T$_{\rm{P}}$deducing the PM part of the FC curve. (b) A comparison between the T$_{\rm{P}}$ values from magnetization and resistivity curves as a function of $x$.}%
\end{figure}

\begin{figure}%
\centering
\subfigure[][]{%
\includegraphics[width=0.5\textwidth]{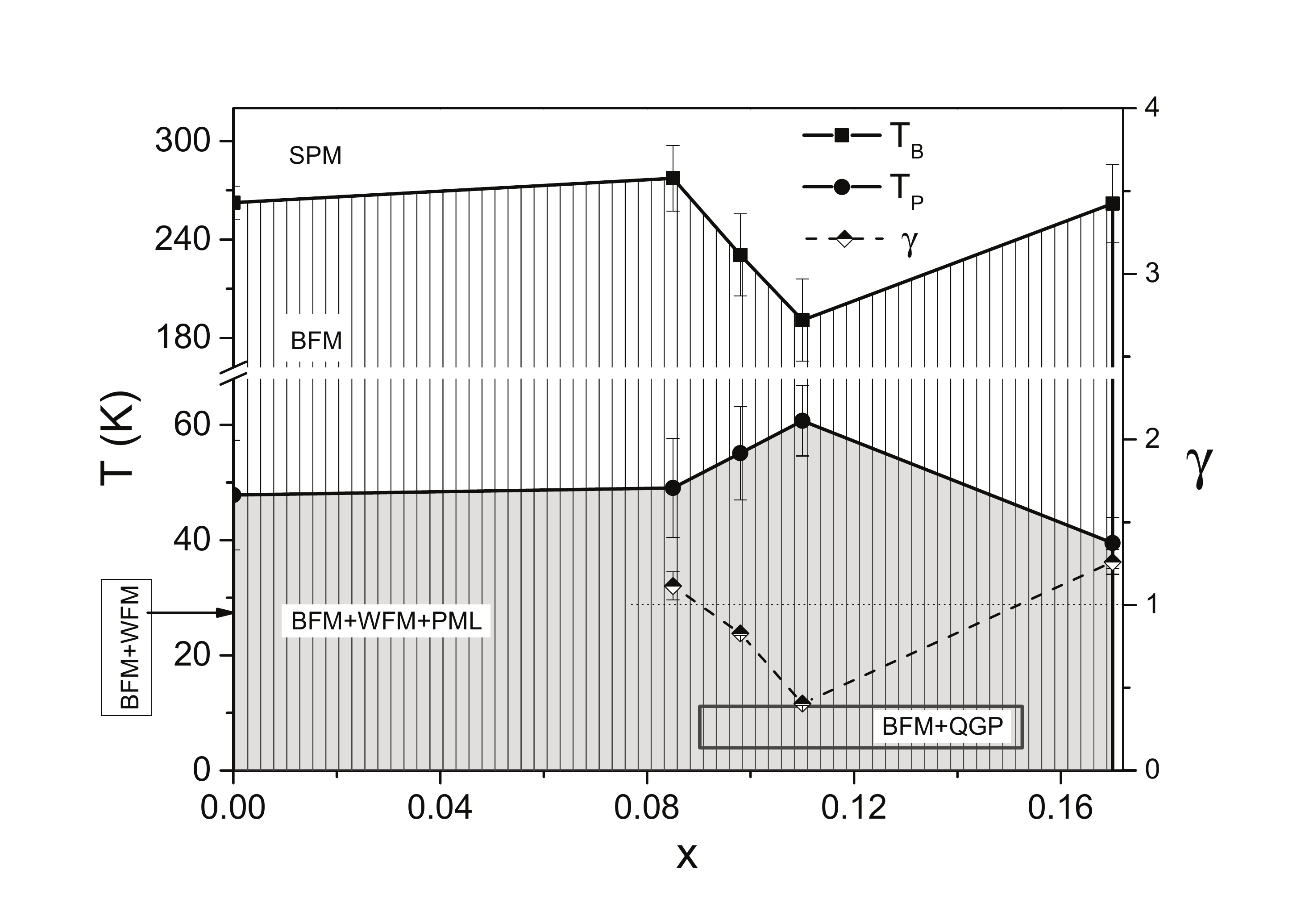}}%
\hspace{8pt}%
\subfigure[][]{%
\includegraphics[width=0.5\textwidth]{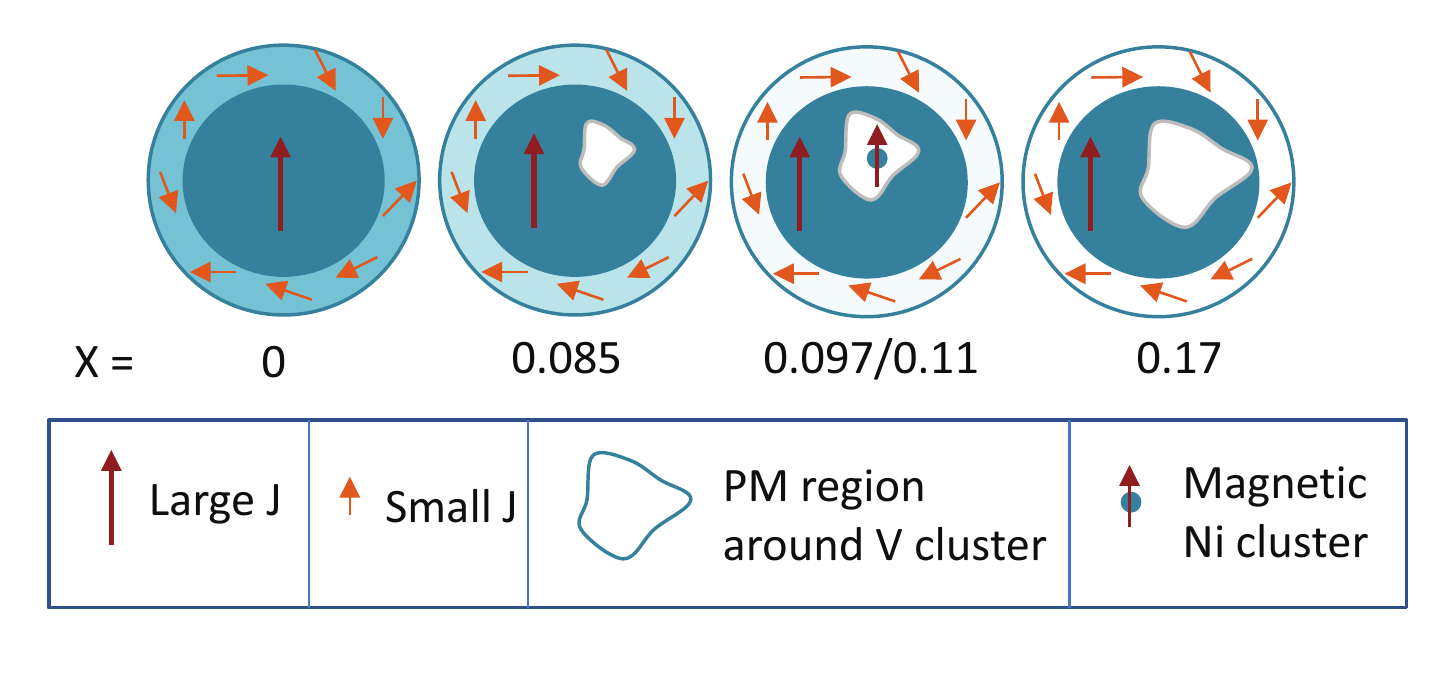}
}%

\caption[]{(a) Magnetic phase diagram of Ni$_{1-x}$V$_x$ nanoparticles. Different regions are represented by: SPM - superparamegntic, BFM - blocked ferromagnetic, WFM - ferromagnetic with weak exchange interaction, PML - paramagnetic-like, and QGP - quantum Griffiths phase.
(b) Schematic pictures of nanoparticle magnetic structures at different concentrations. Large $\mathsf{J}$ - spins with large FM exchange interaction giving rise to a single large moment, small $\mathsf{J}$ - spins with weak FM exchange interaction. Lighter shade implies weaker exchange interaction. The concentrations $x$  = 0.097 and 0.011 contain QGP's.}%
\end{figure}

\begin{figure}%
\centering
\includegraphics[width=0.50\textwidth]{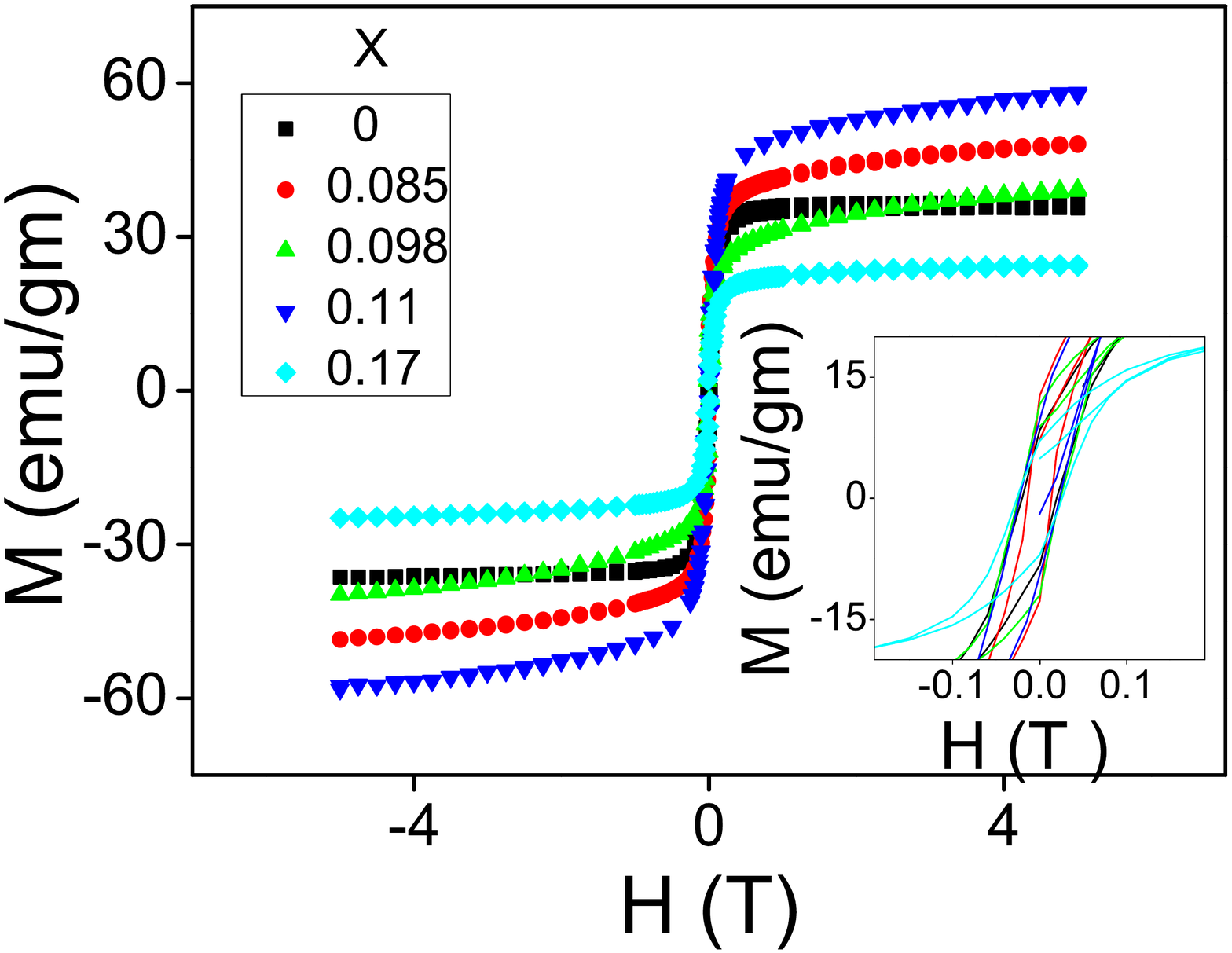}%
\hspace{8pt}%

\caption[]{  M-H curves at 2 K for $x$ = 0.000, 0.085, 0.098, 0.17 and at 5 K for $x$ = 0.11. Inset: Low-field region of the M-H curves showing hystereses.}%
\end{figure}

On introducing the first increment of V in Ni ($x$ = 0.085 in this report), the V atoms statistically occupy Ni sites either as single-atom impurities or as clusters. According to Friedel,\cite{Friedel} a V impurity creates a spin reduction on the neighbouring Ni sites. The effect of V substitution on the surface would then be to reduce $\mathsf{J}$ further and make the surface essentially PM. The same process happens in the bulk. Additionally, the V clusters are proposed to create isolated PM zones in the bulk, as shown schematically in Fig. 9. In this scenario, the low temperature susceptibility, which does not saturate till the lowest measured temperature (5 K), must follow either a Curie-Weiss law $\chi = \rm {C/(T - T_C)^ {-\gamma}}$ (1.25 $\leq \gamma \leq$ 1.3) of ferromagnetism with T$_{\rm C} <$ 5 K, or the Curie law $\chi \propto$ T$^{-\gamma} $ ($\gamma$ = 1) of paramagnetism.\cite{Ashcroft} Assuming T$_{\rm C} \sim$ 0 in the former case, $\gamma$ can be obtained from the slope of the log $\chi$ versus log T curve at the lowest temperatures and must lie between 1 and 1.33. The curve and its linear fit are shown in Fig. 11. The $\gamma$ value is 1.12, which is in agreement with the model of the pure Ni nanoalloy as discussed above.

A further increase in V concentration, then, is supposed to make the nanoalloys more and more PM and hence one should get the $\gamma$ value in the low temperature region $\sim$ 1 for all other compositions. However, and as can be seen from the Fig. 11, $\gamma'$s obtained this way for $x$ = 0.098 and 0.11 are 0.83 and 0.41, respectively, deviating considerably and clearly from the universal power-law (1 $\leq \gamma \leq$ 1.33) behaviour. This is characteristic of a QGP, wherein a PM phase possesses a magnetic cluster inclusion, as reported by Ubaid-Kassis {\it et al}. for bulk Ni$_{1-x}$V$_x$ alloys.\cite{Schroeder10} The more the $x$ deviates from $x_c$ in the PM region, the less the value of $\gamma$ is and the stronger the QGP nature becomes.\cite{Schroeder10} In the studied nanoparticles, an increased V concentration would result in bigger PM zones in the vicinity of bigger V clusters. Statistically, there is a finite probability of much smaller Ni clusters to be enclosed within these larger PM clusters, as shown schematically in the Fig. 9, giving rise to the observed quantum Griffiths behaviour. On augmenting the V content further, the PM zone would expand. The simultaneous depletion of Ni content would then reduce the possibility of Ni clusters to get included in the PM zone and the nanoparticle would now comprise only of the PM zones in addition to the rest of the FM volume. A schematic of this structure is also shown in Fig. 9. This way, one would expect the $\gamma$ value to enhance back to $>$ 1. This is indeed the case for $x$ = 0.17 with $\gamma$ = 1.23, as is apparent from the Fig. 11. Figure 9 also includes a plot between $\gamma$ and $x$, and the region in the range $0.09 \leq x \leq 0.14$ having a signature of QGP in the phase diagram is identified.

\begin{figure}%
\centering
\includegraphics[width=0.6\textwidth]{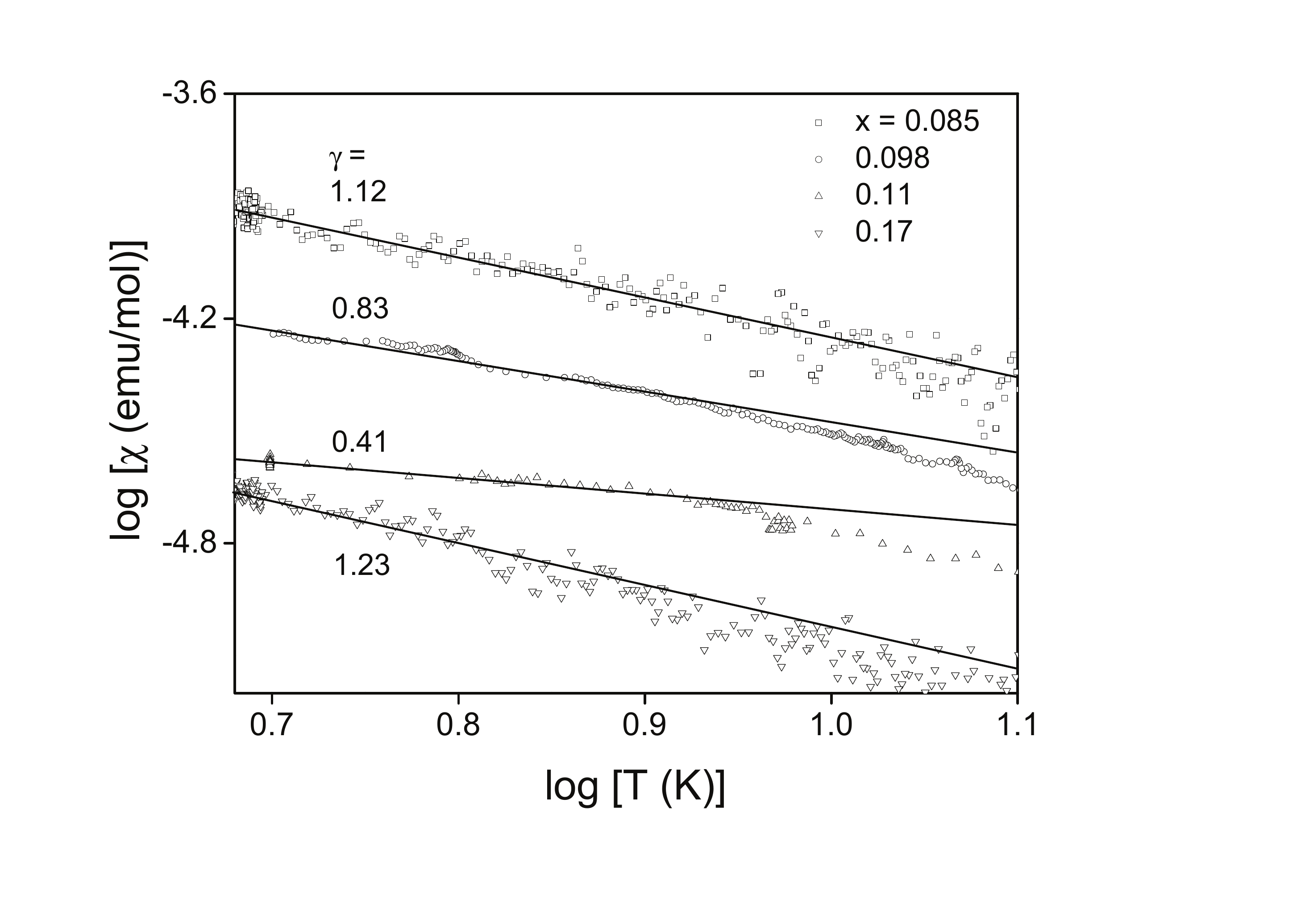}%
\caption[]{Log $\chi$ versus log T curves (symbols) at low T range for various $x$ values and their linear fits (solid lines). Shown also is the $\gamma$ value for each curve.}%
\end{figure}

\section{Conclusion}

Ni$_{1-x}$V$_x$ (0 $\leq$ x $\leq$ 0.17) nanoalloys of mean diameters 18 - 33 nm were prepared by a chemical reflux method with hydrazine hydrate as the reducing agent and diethanolamine as the surfactant. The compositions of the finally synthesized nanoalloys were determined using EDAX. The particle sizes were calculated from the FESEM images, while HRTEM images and SAED patterns displayed magnetic and crystalline structure of the nanoparticles, respectively. The magnetic nature was later confirmed from M-H and M-T measurements. Further, XRD and XPS spectra confirmed that the nanoalloys were indeed a solid solution of Ni and V without any trace of oxides. The temperature dependence of resistivity, apart from ascertaining the metallic nature of the nanoparticles, revealed a PM phase coexisting with the FM phase in the nanoparticles in the form of a peak in 40 K - 60 K temperature range. The M-T curves of all the samples exhibited SPM nature of the nanoparticles. However, each of these curves was associated with a PM-like increase below an $x$-dependent temperature T$_{\rm P}$, coinciding roughly with the peak position in the corresponding resistivity curve. While on the one hand this increase suggests the coexistence of a weak FM phase with the blocked FM below T$_{\rm P}$ in the case of pure Ni nanoparticles, it is found to be associated with a PM-like phase in the nanoalloys, coexisting with the blocked FM phase. The weak FM phase is explained with the existence of weakly interacting surface spins. A fitting of the low-T log $\chi$ versus log T data suggests that while the PM-like phase is really paramagnetic in nature associated with a universal power-law behaviour with the exponent $\gamma$ in the range 1 $\leq \gamma \leq$ 1.33 for $x$ = 0.085 and 0.17, the exponent is found to be non-universal ($<$ 1) for $x$ = 0.097 and 0.11. The non-universal power-law is characteristic of a QGP. A T - $x$ phase diagram has been drawn to show the various existing phases, including the QGP, in the Ni$_{1-x}$V$_x$ nanoalloys.

\vspace{10 mm}
{\bf ACKNOWLEDGEMENTS}
\vspace{10 mm}

 P. Swain acknowledges the financial support from the Council of Scientific and Industrial Research (CSIR), New Delhi.


\vspace{5 mm}
\begin{thebibliography}{99}
\bibitem{Nicklas99}M. Nicklas, M. Brando, G. Knebel, F. Mayr, W. Trinkl, and  A. Loidl, Phys. Rev. Lett. {\bf
82}, 4268 (1999).
\bibitem{Ododo77}J. C. Ododo, and B. R. Coles, Journal of Physics F: Metal
Phys. {\bf 7}, 11 (1977).
\bibitem{Muellner} W. C. Muellner and J. S. Kouvel, Phys. Rev. B {\bf 11}, 4552 (1975).
\bibitem{Schroeder10}S. Ubaid-Kassis,  T. Vojta, and A. Schroeder, Phys. Rev. Lett. {\bf 104}, 066402 (2010).
\bibitem{subir} Subir Sachdev, {\it {Quantum Phase Transitions}} (Cambridge Unversity Press, Cambridge, England, 2011).
\bibitem{Schroeder11}S. Ubaid-Kassis,  T. Vojta, and A. Schroeder, J. Phys.: Condens. Matter {\bf 23}, 094205 (2011).
\bibitem{sachdev11} S. Sachdev and B. Keimer, Phys. Today {\bf 64}(2), 29 (2011).
\bibitem{weismann} A. Weismann, M. Wenderoth, S. Lounis, P. Zahn, N. Quaas, R. G. Ulbrich, P. H. Dederichs, and S. Bl{\"u}gel, Science {\bf 323}, 1190 (2009).
\bibitem{swain15}P. Swain, Suneel K. Srivastava, and Sanjeev K. Srivastava, Phys. Rev. B {\bf 91}, 045401 (2015).
\bibitem{Hiroi} K. Hiroi, H. Kura, T. Ogawa, M. Takahashi and T. Sato, J. Phys.: Condens. Matter {\bf 26}, 176001 (2014).
\bibitem{Kazan} T. G. Altincekic, I Boz, A. C. Basaran, B. Aktas and S. Kazan, J. Supercond. Nov. Magn. {\bf 25} 2771 (2012).
\bibitem{He} X. He, W. Zhong, C-T. Au and Y. Du, Nanoscale Res. Lett. {\bf 8}, 446 (2013).
\bibitem{Fonesca} F. C. Fonesca, G. F. Goya, R. F. Jardim, R. Muccillo, N. L. V. Carreno, E. Longo and E. R. Leite, Phys. Rev. B {\bf 66}, 104406 (2002).
\bibitem{Senapati} S. Senapati, S. K. Srivastava and S. B. Singh, Nanoscale {\bf 4}, 6604 (2012).
\bibitem{Bambhaniya13}K. G. Bambhaniya, G. S. Grewal, V. Shrinet, N. L. Singh, and T. P. Govindan, International Journal of Green Energy. {\bf 10},640-646 (2013).
\bibitem{ABC} Something to show that magnetic particles agglomerate in HRTEM.
\bibitem{Qi} W. H. Qi and M. P. Wang, J. Nanopart. Res. {\bf 7}, 51 (2005).
\bibitem{PhaseD} resource.npl.co.uk/mtdata/phdiagrams/niv.htm.
\bibitem{Guo} F. Q. Guo and K. Lu, Nanostruct. Mater. {\bf 7}, 509 (1996), and references therein.
\bibitem{Steiner1981}P. Steiner, and S. Huefner, Solid State Commun. {\bf 37}, 79 (1981).
\bibitem{Diep2011}T. Diep, Y. Magnin,  and D. Hoang, ACTA PHYSICA POLONICA A {\bf 121}, 5-6 (2011).
\bibitem{Magnin2011}Y. Magnin, Danh-Tai Hoang, and H. T. Diep {\it arXiv:1101.5789v1} [cond-matter.str-el] 30 jan (2011).
\bibitem{castrillon} M. Castrillon, A. Mayoral, C. Magen, J. G. Meier, C. Marquina, S. Irusta, and J. Santamaria, Nanotechnology {\bf 23}, 085601 (2012).
\bibitem{Shim} H. Shim, P. Dutta, M. S. Seehra, and J. Bonevich, Solid State Comm. {\bf 145}, 192 (2008).
\bibitem{Lueker} A. Leuker at http://www.arne-lueker.de/Objects/work/Magnetic/nanoparticles.html
\bibitem{Blundell} S. Blundell, Magnetism in condensed matter, Oxford University Press, N.Y., 2001.
\bibitem{Qin2014} W. Qin, X. Li, Y. Xie, and Z. Zhang, Phys. Rev. B {\bf 90}, 224416 (2014).
\bibitem{Friedel} J. Friedel, Nuovo Cimento {\bf 7}, 287 (1958).
\bibitem{Ashcroft} N. W. Ashcroft and N. D. Mermin, Solid State Physics, Cengage Learning, 1976.

\end{thebibliography}
\end{document}